\documentclass{jfm-final}
\usepackage{graphicx}
\usepackage{epstopdf, epsfig}

\shorttitle{Remote recoil and the Aharonov--Bohm effect}
\shortauthor{Michael Edgeworth McIntyre}

\title{Wave--vortex interactions, remote recoil, the \ab\
              effect and the Craik--Leibovich equation}

\author{Michael Edgeworth McIntyre\aff{1}
  \corresp{\email{mem@damtp.cam.ac.uk}},
}

\affiliation{\aff{1}Department of Applied Mathematics and
    Theoretical Physics, Cambridge CB3 0WA, UK}

\newcommand{\qv}{\textit{q.\sliver v.}}

\newcommand{\lhs}{left-hand side}
\newcommand{\rhs}{right-hand side}
\newcommand{\te}{thought-experiment}
\newcommand{\wvi}{wave--vortex interaction}
\newcommand{\qg}{quasigeostrophic}
\newcommand{\psm}{pseu\-do\-momentum}
\newcommand{\Psm}{Pseu\-do\-momentum}

\newcommand{\cleq}{Craik--Leibo\-vich equation}
\newcommand{\clvf}{Craik--Leibo\-vich vortex force}
\newcommand{\ab}{Aharonov--Bohm}
\newcommand{\nil}{noninterchangeability of limits}
\newcommand{\tpj}{topological phase jump}
\newcommand{\abpj}{Aharonov--Bohm phase jump}

\newcommand{\oa}{$O(\amp)$}
\newcommand{\oaa}{$O(\amp^2)$}
\newcommand{\oaaaa}{$O(\amp^4)$}
\newcommand{\oeps}{$O(\frma)$}
\newcommand{\oee}{$O(\frma^2)$}

\newcommand{\oaaeo}{$O(\amp^2\frma^0)$}
\newcommand{\oaae}{$O(\amp^2\frma^1)$}
\newcommand{\oaoee}{$O(\amp^0\frma^2)$}
\newcommand{\oaaee}{$O(\amp^2\frma^2)$}
\newcommand{\oaaeee}{$O(\amp^2\frma^3)$}
\newcommand{\oaaaaeo}{$O(\amp^4\frma^0)$}
\newcommand{\oaaaae}{$O(\amp^4\frma^1)$}

\newcommand{\sliver}{{\hskip0.6pt}}

\newcommand{\Sliver}{{\hskip1.2pt}}
\newcommand{\antisliver}{{\hskip-0.6pt}}

\newcommand{\Antisliver}{{\hskip-1.2pt}}
\newcommand{\sgn}{\mbox{\rm sgn}}
\newcommand{\imag}{\mbox{\rm i}}
\newcommand{\km}{{\Sliver km}}
\newcommand{\degree}{{$^\circ$}}

\newcommand{\half}{\tfrac{1}{2}}
\newcommand{\tinyhalf}{{\tiny\frac{1}{2}}}
\newcommand{\quarter}{\tfrac{1}{4}}

\newcommand{\amp}{a}
\newcommand{\wspeed}{c}
\newcommand{\vspeed}{U}

\newcommand{\butr}{\bu_{\rm tr}}
\newcommand{\ktyp}{k}
\newcommand{\envel}{A}
\newcommand{\radius}{r}
\newcommand{\Radius}{r'}

\newcommand{\xextent}{X}

\newcommand{\length}{L}
\newcommand{\width}{W}
\newcommand{\wakewidth}{w}
\newcommand{\bk}{\boldsymbol{k}}
\newcommand{\bC}{\boldsymbol{C}}
\newcommand{\bCabs}{\boldsymbol{C}^{\rm abs}}
\newcommand{\Cabs}{C^{\rm abs}}
\newcommand{\bxi}{\boldsymbol{\xi}}
\newcommand{\bu}{\boldsymbol{u}}

\newcommand{\bx}{\boldsymbol{x}}
\newcommand{\bxvec}{\boldsymbol{\hat x}}
\newcommand{\byvec}{\boldsymbol{\hat y}}
\newcommand{\bzvec}{\boldsymbol{\hat z}}
\newcommand{\bthetavec}{\raisebox{-0.6pt}{$\boldsymbol{\hat{\theta}}$}}
\newcommand{\bnvec}{\boldsymbol{\hat n}}
\newcommand{\fcoriolis}{f}
\newcommand{\bfcoriolis}{\boldsymbol{f}}
\newcommand{\bpmom}{\hbox{\textbf{\textsf p}}}
\newcommand{\bpmomint}{\hbox{\textbf{\textsf P}}}
\newcommand{\pmom}{\hbox{\textsf p}}

 \newcommand{\bfpmom}{\hbox{\textbf{\textsf B}}}
\newcommand{\bfpmomo}{\hbox{\textbf{\textsf B}}_{\rm o}}
\newcommand{\bfpmomw}{\hbox{\textbf{\textsf B}}_{\rm w}}
 \newcommand{\fpmom}{\hbox{\textsf B}}

\newcommand{\bimp}{\boldsymbol{I}}
\newcommand{\bubars}{\overline{\boldsymbol{u}}^{\rm S}}
\newcommand{\us}{\overline{u}^{\rm S}}

\newcommand{\bubarl}{\overline{\boldsymbol{u}}^{\rm L}}
\newcommand{\bubarlh}{\overline{\boldsymbol{u}}_{\mbox{\rm\scriptsize H}}^{\rm L}}
\newcommand{\bubarlb}{\overline{\boldsymbol{u}}_{\mbox{\rm\scriptsize B}}^{\rm L}}
\newcommand{\ubarlh}{\overline{u}_{\mbox{\rm\scriptsize H}}^{\rm L}}
\newcommand{\vbarlh}{\overline{v}_{\mbox{\rm\scriptsize H}}^{\rm L}}
\newcommand{\ubarl}{\overline{u}^{\rm L}}

\newcommand{\wbarl}{\overline{w}^{\rm L}}

\newcommand{\bubar}{\overline{\boldsymbol{u}}}

\newcommand{\bF}{\boldsymbol{F}}
\newcommand{\bR}{{\antisliver\boldsymbol{R}\sliver}}
 \newcommand{\nbR}{{\antisliver{R}\sliver}}
\newcommand{\nbRo}{{\antisliver{R}_{\rm o}\sliver}}
\newcommand{\nbRw}{{\antisliver{R}_{\rm w}\sliver}}
\newcommand{\bomega}{\boldsymbol{\omega}}
\newcommand{\bomegabar}{\overline{\boldsymbol{\omega}}}
\newcommand{\bomegatilde}{{\boldsymbol{\widetilde\omega}}}

\newcommand{\omegatilde}{{\widetilde\omega}}
\newcommand{\omegatildeb}{{\widetilde\omega}_{\mbox{\rm\scriptsize B}}} 
\newcommand{\bomegatildeh}{{\boldsymbol{\widetilde\omega}}_{\mbox{\rm\scriptsize H}}}
\newcommand{\nablasq}{{\nabla}^2}
\newcommand{\bnablah}{\boldsymbol{\nabla}_{\!\mbox{\rm\scriptsize
H}}^{\phantom{L}}}
\newcommand{\nablahsq}{\nabla_{\!\mbox{\rm\scriptsize H}}^2}
\newcommand{\cross}{\!\times\antisliver}
\newcommand{\cdott}{\!\cdot}
\newcommand{\Dbarl}{\overline{D}^{\rm L}}
\newcommand{\Dbarlh}{\overline{D}_{\antisliver\mbox{\rm\scriptsize H}}^{\rm L}}

\newcommand{\psitilde}{\tilde\psi}

\newcommand{\psitildeb}{\tilde\psi_{\mbox{\rm\scriptsize B}}^{\phantom{i}}}

\newcommand{\LD}{L_{\rm D}}
\newcommand{\htilde}{\tilde h}
\newcommand{\rhotilde}{\tilde\rho}
\newcommand{\pbarl}{\overline{p}^{\rm L}}
\newcommand{\qtilde}{\tilde{q}}
\newcommand{\Ro}{{\rm Ro}}

\newcommand{\frma}{\epsilon}
\newcommand{\errorpsi}{\delta\psitilde}
\newcommand{\thetatilde}{\tilde\theta}
\newcommand{\lind}{l}
\newcommand{\mind}{m}

\usepackage{amssymb}
\usepackage{amsmath}

\begin{document}

\maketitle

\begin{abstract}
Three
examples of
non-dissipative yet cumulative interaction between a single wavetrain
and a single vortex are analysed,
with a focus on effective recoil forces, local and remote.
Local recoil occurs when the wavetrain overlaps the vortex core.
All three
examples comply with the \psm\ rule.
The first two
examples are two-dimensional and
non-rotating (shallow~water or gas dynamical).
The third
is rotating, with deep-water gravity waves inducing an
Ursell
``anti-Stokes flow''.
The~Froude or Mach number, and the Rossby number in the third
example, are assumed small.
Remote recoil is all or part of the interaction in
all three
examples, except in one special limiting case.
That case is found only within
a severely restricted parameter regime and is
the only case in which, exceptionally,
the effective recoil force can be regarded as purely local
and identifiable with the celebrated \clvf\ --
which corresponds, in the
quantum fluids literature, to the Iordanskii force due to a
phonon current incident on a vortex.
Another peculiarity of that exceptional case
is that the only significant wave refraction effect is the
\ab\ \tpj.
\end{abstract}

\vspace{-0.7cm}

\section{Introduction}
\label{sec:intro}

In the vast literature on wave--mean and wave--vortex interactions, there
is a tradition of thinking in terms of wave-induced mean forces and
the associated wave-induced momentum fluxes or radiation stresses.  The
tradition goes back many years, to the work of Lord Rayleigh, L\'eon Brillouin
and other great physicists.  It continues today in, for instance, work on
the fluid dynamics of atmospheres and oceans, as well as on quantum vortices
where the wave-induced mean forces are called ``Iordanskii forces''.

Within the atmosphere--ocean community, the force-oriented viewpoint is
important because wave-induced mean forces are recognized as key
to solving what used to be three great enigmas -- three
grand challenges -- in atmospheric science.  They were to understand the
quasi-biennial oscillation of the zonal winds in the equatorial stratosphere,
the ``antifrictional'' self-sharpening of jet streams, and the
gyroscopic or Coriolis pumping of global-scale mean circulations in the
stratosphere and mesosphere (i.e.\ between altitudes $\sim$10--100\km)
and the consequent water vapour, ozone and pollutant transport and,
most dramatically, the refrigeration of the summer mesopause --
down to temperatures $\sim 100$\degree C below radiatively determined
temperatures.  The history is tortuous and
goes back to the 1960s, when all three phenomena were observationally
conspicuous but, in terms of mechanism, completely mysterious.
See for instance \citet{Wallace:1968},
\citet{Fritts:1984}, \citet{Holton:1995}, \citet{Baldwin:2001},
\citet{Dritschel:2008}, and \citet{Garcia:2017}.

Recognition of wave-induced mean forces as key to solving all three enigmas,
and as essential components of weather and climate models, constituted a
gradual, but major, paradigm shift regarding the nature of large-scale
momentum transport in atmospheres and oceans.  Before the 1960s,
such transport tended to be thought of
in terms of turbulent eddy viscosities, missing the point that
wave-induced momentum transport can be a long-range process more likely to
dominate, as in fact it does, over large scales limited not by
parcel displacements and mixing lengths but instead by the far greater distances
over which waves can propagate.

Much of the atmosphere--ocean literature, especially where it deals with
mean forces induced by gravity waves, often takes for granted that the
forces can be computed from linearized wave theory alone using
what is now called the \emph{\psm\ rule}
\citep[e.g.][hereafter B14]{Buhler:2014}.
That is the accepted basis of gravity-wave ``parametrization schemes''
in weather and climate models, designed to incorporate the mean forces
coming from gravity waves whose wavelengths are too small to be resolved
explicitly.  It is also the usual basis on which, for instance, Iordanskii
forces are computed \citep[e.g.][]{Sonin:1997,Stone:2000a}.  \Psm, also
called quasimomentum or wave momentum, or phonon momentum, is the
linear-theoretic wave property whose nondissipative conservation depends,
through Noether's theorem, on translational invariance of the mean or background
state on which the waves propagate, as distinct from translational
invariance of the entire physical system, background plus waves, which
implies conservation of momentum.

In a linearized ray-theoretic
description the \psm\ $\bpmom$ per unit mass is ${\cal A}\sliver\bk$,
where $\bk$
is the wavenumber vector and $\cal A$ the wave-action, i.e.\
wave-energy divided by intrinsic frequency, per unit mass
\citep[e.g.][]{Bretherton:1968}.  The
wave-action, wave-energy and \psm\ are linear-theoretic wave properties and
are \oaa\ in magnitude, where $\amp$ measures wave amplitude
(and will be defined in such a way that
$\amp \ll 1$ validates linearization).
The \psm\ rule says that \oaa\ wave-induced mean forces can be
calculated as if \psm\ were momentum and the fluid medium were absent.

As discussed for instance in B14, the rule has been justified mainly
for simplified mean flows that are themselves
translationally
invariant.
In such cases it is typical, as is well known,
for persistent mean forces with cumulative effects
to arise only when the waves break or are otherwise dissipated,
leading to a persistent \psm-flux convergence.  However, when the waves
are refracted by realistic, three-dimensional backgrounds involving vortices,
the situation is fundamentally different.  One can get persistent mean
forces with cumulative effects in the absence of wave dissipation.  Also,
it is unclear whether, or when, or in what sense the \psm\ rule should
hold.
The fluid medium is not absent, and
it supports a mean pressure field that mediates
long-range mean forces of the same order, \oaa, as those computed from the
\psm\ rule.  Such pressure fields are not wave properties and
cannot be computed from linearized wave theory alone.  Rather, they require
the solution of equations governing the mean or background state
correct to \oaa.  Cases in which the rule fails for this reason
have long been known, going back as far as Brillouin's
pioneering work on acoustic radiation stress
\citep[e.g.][B14 \S12.2.2]{Brillouin:1936}.

\begin{figure}
\centering
\vspace{0.4cm}
\hspace{1cm}
\includegraphics[width=11 cm, viewport=0 240 800 510,clip]{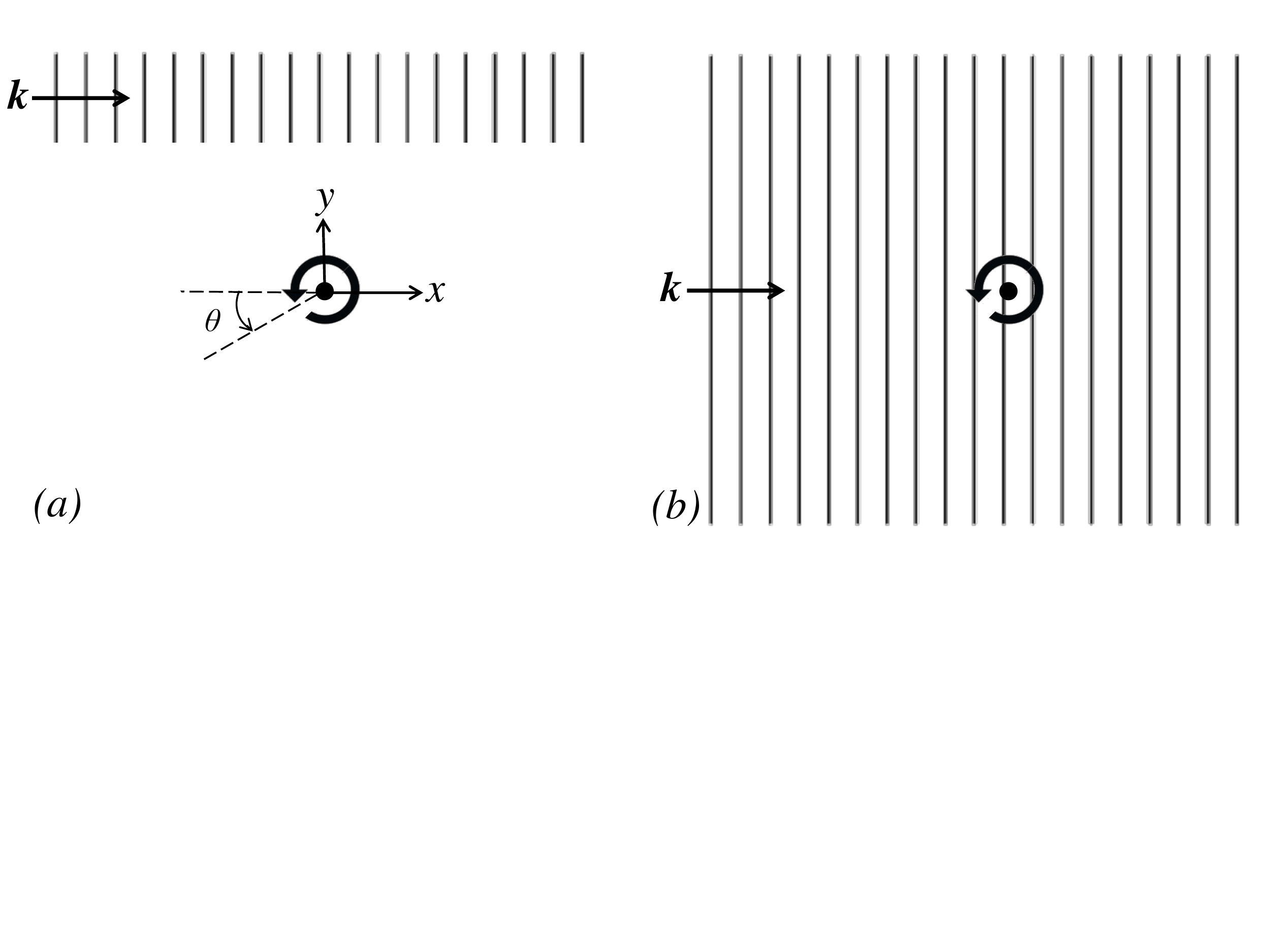}
\vspace{0.1cm}
\caption{Panels~(\textbf{a})
and~(\textbf{b}) are schematics
of \wvi\ problems (i) and~(ii) respectively.
Waves of wavenumber $\bk$ are incident from the left and
are weakly refracted by the vortex.  Rays are
nearly parallel to the $x$~axis.
The azimuthal angle $\theta$ is defined
unconventionally but in a way that will be convenient when
discussing the \ab\ effect.
}
\vspace{-0.35cm}
\label{fig:probs1-2}
\end{figure}

For gravity-wave parametrization, in particular, there are therefore
unresolved questions as to how to compute, and indeed how to think about,
the wave-induced mean forces for realistic, three-dimensional backgrounds.
Current parametrization schemes ignore these questions because they
altogether neglect horizontal refraction, giving rise to what is sometimes
called the ``missing forces'' problem for such schemes.

The simplest wave--vortex problems in which these questions arise are to be
found in a two-dimensional, non-rotating shallow-water or acoustical setting,
with no viscous or other wave dissipation.  Two basic examples, the main
examples to be studied in this paper, are sketched in
figures~\ref{fig:probs1-2}\textit{(a)} and~\ref{fig:probs1-2}\textit{(b)}.
They will be
referred to as problem~(i) and problem~(ii), respectively.
The background flow
is a single vortex whose vorticity is confined to a core of radius
$\radius = \radius_0$, say, with irrotational flow outside.
The coordinates are as shown in figure~\ref{fig:probs1-2}\textit{(a)},
with $\radius^2=x^2+y^2$.  The vortex weakly refracts a train of
gravity waves or sound waves incident from the left.
The refraction induces a small difference
between incoming and outgoing \psm\ fluxes -- corresponding to the
background not being translationally invariant -- and the \psm\ rule leads
us to expect a persistent \oaa\ mean recoil force to be exerted.
That expectation
is independent of whether or not the waves overlap the vortex core.
One reason for studying the two problems side by side is
a desire to understand how overlap or non-overlap affect the
way in which the recoil force arises, and where it is exerted,
as well as whether it complies with the \psm\ rule.

Problem~(i), with no overlap,
has already been studied
in an earlier
paper \citep[][hereafter BM03]{Buhler:2003} but will be revisited here in
order to compare it with problem~(ii), for which new results will be obtained.
Also new will be results for a rapidly-rotating version of problem~(i),
to be defined below and to be referred to as problem~(iii).

Implicit here, as above, is the assumption that the waves can be described
by linearized theory for $\amp \ll 1$ on a background flow of much greater
magnitude.  Our aim is to obtain precise results by analytical means, in
order to gain insight into the questions just noted.  To get analytically
tractable, precisely soluble problems it turns out
that we must also assume, as was done in BM03,
that the background flow and the resulting refraction are very weak
in the sense that
the vortex must be assumed to have small Froude or Mach
number

\vspace{-0.85cm}

\begin{equation}
\frma \:=\:  \vspeed/\wspeed_0  \:\ll\: 1
~,
\label{eq:frma-def}
\end{equation}
where $U$ is a vortex flow speed and $\wspeed_0$
an intrinsic wave speed. \
Thus the analyses to be presented fall within the asymptotic regime
$\amp \:\ll\: \frma \:\ll\: 1$. \
For definiteness, $\vspeed$ will be taken to be the flow speed at the
edge $\radius = \radius_0$ of the vortex core, and $\wspeed_0$
the wave speed far from the core.
For general $\radius \geqslant \radius_0$
the wave speed
$\wspeed = \wspeed(r) = \wspeed_0\{1 + O(\frma^2\radius_0^2/\radius^2)\}$,
from the Bernoulli effect and the $\radius^{-1}$ dependence of the
vortex flow speed.

The regime $\amp \:\ll\: \frma \:\ll\: 1$ also encompasses the
celebrated Lighthill theory of spontaneous sound emission from,
and scattering by,
unsteady systems of vortices.
It can be contrasted with, for instance, the regime
$\amp \:\sim\: \frma \:\ll\: 1$
\citep[e.g.][]{Lelong:1991,Ward:2010,Thomas:2017},
in which
wave--vortex interactions
of the resonant-triad type are possible.  The vortical field,
if sufficiently complex spatially,
can then act as a passive ``catalyst'' of wave--wave energy transfer
very like the Bragg scattering or ``elastic scattering'' studied in
\citet{McComas:1977b}, in a somewhat different context.
Yet another regime of interest, one that has
been studied very often in past decades, is
$\amp^2 \:\sim\: \frma \:\ll\: 1$, for instance in connection with the
generation of Langmuir vortices by the nondissipative Craik--Leibovich
instability
\citep[e.g.][B14 \S11.3]{Craik:1976,Leibovich:1980}.
Indeed the regime $\amp^2 \:\sim\: \frma \:\ll\: 1$
arises naturally in a great variety of problems where
mean flows are generated nondissipatively, from rest,
entirely through the presence of waves.
Then refraction of the waves by the mean flow comes in only at higher
order.  Further such examples include, among many others, those studied by
\citet{LonguetHiggins:1964b}, \citet{Bretherton:1969},
\citeauthor{McIntyre:1981}
(\citeyear{McIntyre:1981}, \citeyear{McIntyre:1988}),
\citet{Wagner:2015}, \citet{Haney:2017},
\citet{Thomas:2018}, and
\citet{Thomas:2019}.

Returning now to problems~(i)--(iii), in which
$\amp \:\ll\: \frma \:\ll\: 1$ and refraction takes place
at leading order in $\frma$,
it will be shown in this paper
that the \psm\ rule is satisfied not only in problem~(i) but
also in the other two problems, to leading order at least.
In all three problems,
the background feels a persistent \oaae\ mean recoil force satisfying the rule.

Problem~(ii) is a classical version of the phonon--vortex
problem studied in the quantum vortex literature.
This classical version
is considered for instance by \citet{Sonin:1997} and \citet{Stone:2000a}, who
take the wavetrain to be infinitely wide, and incident from $x=-\infty$.
They argue not only that the \psm\ rule holds but also that
there is a remarkable simplification,
namely that the dominant wave-refraction effect,
the sole effect that comes in at leading order, \oaae, is the topological
phase jump arising from what is called \ab\ effect.  We will find, however
-- with cross-checks from two independent methods --
that another, quite different refraction effect is
also relevant in problem~(ii),
except in one special limiting case where the \abpj\ is
indeed dominant.  In that special case the length of the wavetrain is taken
to infinity first, followed by the width.  If the order of limits is
reversed, a different answer is obtained and
the \abpj, while still relevant,
is no longer the only relevant refraction effect.
However, as already emphasized we find that the \psm\ rule is still
satisfied to leading order, that is, correct to \oaae.

The
\abpj\ is a topological property of the wave~field
most simply expressed
via the following far-field solution to the linearized equations.
The solution is well~known in the quantum literature and will
be verified below,
in
\S\ref{sec:eqns} and Appendix~\ref{sec:appa}.
For sufficiently large $\radius$, and outside
a relatively narrow ``wake'' region surrounding the positive $x$~axis,
the wave~field has the asymptotic form
$\envel\exp(\imag\sliver\Phi)$ where the amplitude
$\envel=$ \oa\ is a real constant,
with error $O(\amp\sliver\frma^2\radius_0^2/\radius^2)$,
and the phase $\Phi$ is given by

\vspace{-0.65cm}

\begin{equation}
\Phi
~=~
k_0(x-\wspeed_0 t) \,-\, \alpha\theta \,+\, \mbox{\rm const.}
\,+\,
 O(\frma^2\radius_0^2/\radius^2)
~.
\label{eq:disloc-wavefield}
\end{equation}
The incident wavenumber $k_0$ is a constant.
The azimuthal angle $\theta$ is defined as
in figure~\ref{fig:probs1-2}\textit{(a)} and
ranges from $-\pi$ to $\pi$, while
$\alpha$ is a constant defined by

\begin{figure}
\centering
\vspace{-5.1cm}
\includegraphics[width=16.3cm, viewport=0 -10 600
590,clip]{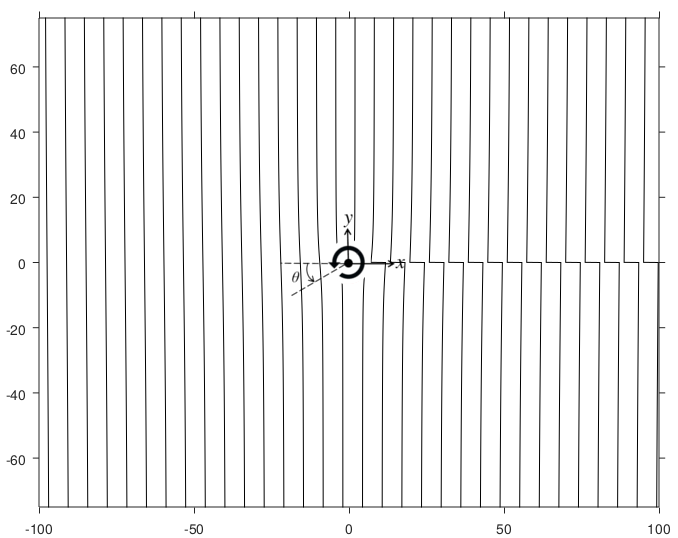}
\vspace{-0.6cm}
\caption{Wavecrests plotted from the far-field solution
(\ref{eq:disloc-wavefield}), with $\alpha=0.75$.
The unit of length is taken as
$k_0^{-1}$ so that the unrefracted wavelength is $2\pi$.
The \abpj\ appears as a
phase discontinuity on the positive $x$-axis.  In a full solution this
discontinuity is smoothed out across a relatively narrow ``wake'' region.
}
\vspace{-0.22cm}
\label{fig:wavecrests-exaggerated}
\end{figure}

\vspace{-0.35cm}

\begin{equation}
\alpha
~=~
\Gamma k_0/2\pi \wspeed_0
~=~
\vspeed k_0\sliver\radius_0/\wspeed_0
~=~
k_0\sliver\radius_0\sliver\frma
\label{eq:alpha-super-small}
\end{equation}
where $\Gamma\sliver$
is the Kelvin circulation of the vortex.
The phase jump $2\pi\alpha$ across the positive $x$ axis is the \abpj,
a topological defect of (\ref{eq:disloc-wavefield}).
In a full solution it is smoothed out across the wake region.
It measures the effect of the vortex flow in compressing
the wavetrain on the positive-$y$ side and stretching it on the negative.
The wavecrest shapes $\Phi =$ const.\ described
by (\ref{eq:disloc-wavefield}), with the error term neglected,
are plotted in figure~\ref{fig:wavecrests-exaggerated}
for $\alpha = 0.75$, fixing the
phase jump
at three-quarters of a wavelength.  Depending on the value of
$k_0\sliver\radius_0$ this can take us outside
the range of validity of
our asymptotic regime (see for instance \S\ref{sec:refrac-within-wake}), but
$\alpha = 0.75$ is chosen here 
to make the refraction effects visible in the figure.
They include the other relevant  effect already mentioned, which is
that, except on the $y$~axis, the wavecrests are slightly rotated away
from the $y$ direction, by $O(\frma\radius_0/\radius)$.
This latter effect is also important in problem~(i), as noted in BM03
and in B14 \S14.2.

There remains the question of \emph{where} the wave-induced recoil force is
exerted.  The question is ambiguous as it stands, and can be answered in
more than one way, but there is one and only one way that avoids bringing
in the \oaa\ mean pressure field.  It has the further advantage of being
the only way that is relevant to gravity-wave parametrization.  It is to ask,
then answer, the question thus: if the waves were removed and the recoil force
exerted artificially, as an external applied force,
where should it be exerted in order to have the same
effect on the mean flow?  As shown in BM03
and B14 \S14.2,
in the case of problem~(i), the answer is not at locations
where the waves are refracted -- as a naive invocation of the \psm\ rule
might suggest -- but, rather, solely at the location of the vortex core.
BM03 therefore called the recoil ``remote''.  In problem~(i),
the vortex core can be arbitrarily distant from locations where the waves
are refracted.  There is of course nothing mysterious about this remoteness
-- no violation of Newton's Third Law -- because
pressure fields
can mediate actions and reactions continuously, over substantial distances,
just as they do in ordinary vortex--vortex interactions.

To arrive at this picture
BM03 relied mainly on a \te\ in which an
artificial ``holding force'' was applied to the vortex core, in such a way
as to cancel the recoil due to the waves.  It was shown by careful analysis
that, by applying this holding force, the mean flow and wave~field
can be kept exactly steady, with exactly constant total momentum.
Here, following B14 \S14.2,
we ask instead
how the mean flow responds to the net \psm\ flux
in the absence of a holding force.
The answer is then that the vortex translates,
and keeps on translating persistently, in a direction perpendicular
to the recoil force -- a classic Magnus-force-like scenario.
It translates because it is advected by an \oaa\ ``Bretherton flow''
induced by the wave~field \citep{Bretherton:1969}.
With no holding force, therefore, the \emph{Kelvin impulse} of the vortex
\citep[e.g.][eq.~(7.3.7), and
             eq.~(\ref{eq:impulse-def}) below]{Batchelor:1967}
changes
in just the same way as if the waves were removed and the recoil force
artificially applied to the vortex core.
Because $\amp^2 \ll \amp \ll \frma$, the wave field
can still be treated as steady.  And
in each problem studied here and in BM03, the Bretherton flow organizes
itself such that the rate of change of impulse corresponds to a
recoil force that is consistent with the \psm\ rule.

\begin{figure}
\centering
\vspace{+0.2cm}
\includegraphics[width=12cm, viewport=0 60 720 450,clip]{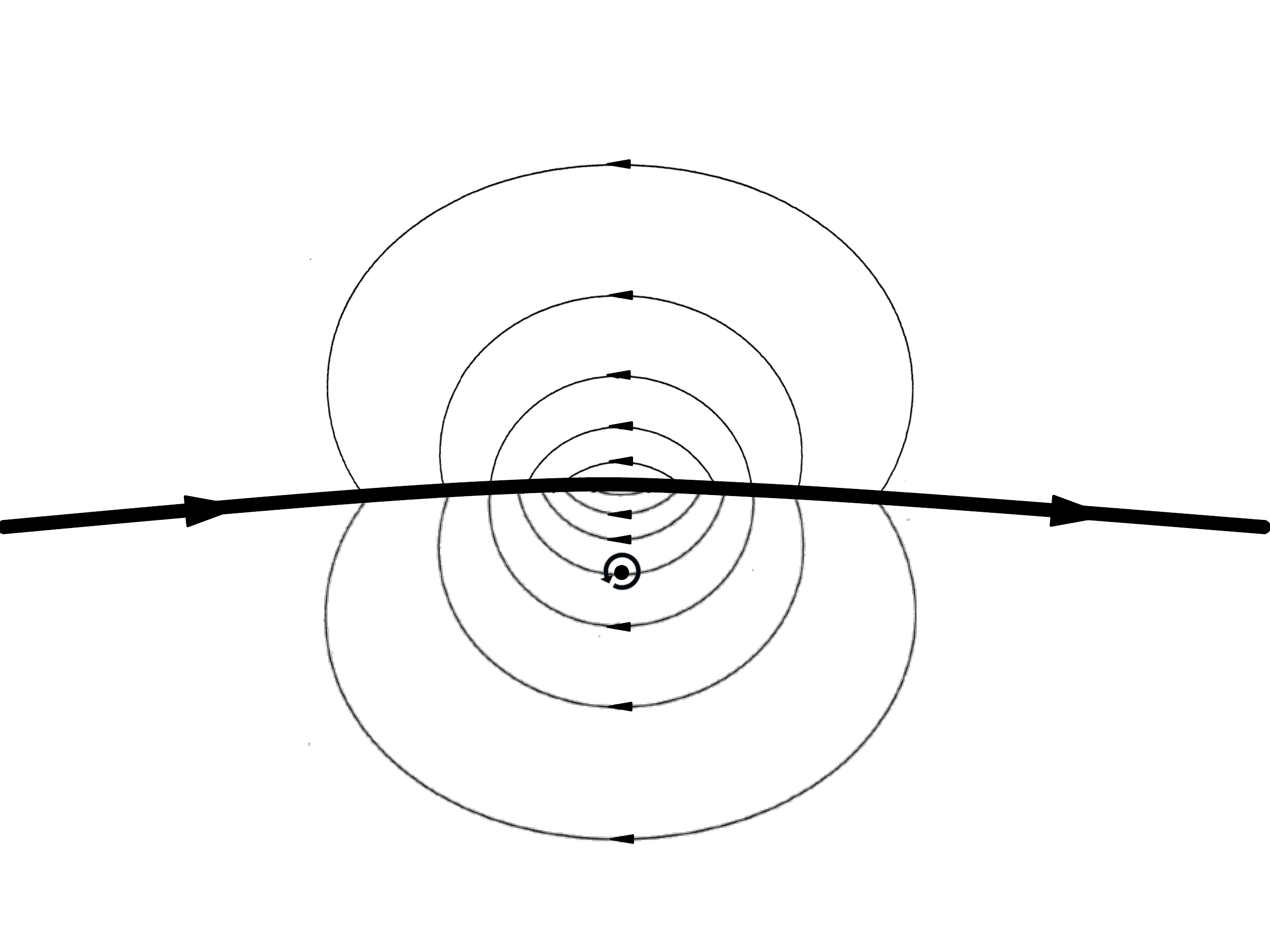}
\caption{Schematic of the Bretherton flow arising in a version of
problem~(i) studied in BM03.
The \oaa\ mean flow within a narrow wavetrain, whose ray path is shown by
the heavy curve, is dominated by the Stokes drift.  A small portion
of its mass flux, \oaae\ in this case,
leaks sideways as a consequence of wave refraction.
To describe this situation the refraction problem must be considered
correct to two orders in \sliver$\frma$\sliver, as was done in \S5.1 of BM03.
Refraction effects enter at both orders, not only the
\oeps\ effects illustrated in figure~\ref{fig:wavecrests-exaggerated} but
also an \oee\ change in the direction of the absolute group velocity,
exaggerated in this schematic.
}
\vspace{-0.35cm}
\label{fig:bretherton}
\end{figure}

By way of illustration,
figure~\ref{fig:bretherton} depicts schematically 
the Bretherton flow in a version of problem~(i)
solved in \S5.1 of BM03, \qv\ for the analytical details.
The heavy curve represents a narrow wavetrain that is slightly deflected as it
goes past the vortex, in such a way that the net \psm\ flux into the region
points in the positive $y$ direction.
Ray theory is used to describe the waves,
as throughout BM03, assuming $k_0\sliver\radius_0 \gg 1$.
The \oaa\ mean flow within the wavetrain
is dominated by the Stokes drift.  A small portion
of its mass flux
leaks sideways as a consequence of wave refraction, forming the Bretherton
flow, which is irrotational
outside the wavetrain.
In the example shown it advects the vortex core leftward,
in the negative $x$ direction.  The corresponding recoil force
-- a force that would move the vortex core leftward in the absence of waves --
is therefore a force
in the positive $y$ direction, like the net \psm\ flux.  Its magnitude
is shown by BM03's analysis to be consistent with the \psm\ rule.

In BM03 and B14, as in the present work, it is assumed that the vortex core size
$\radius = \radius_0$ is small enough to allow the core to be carried
bodily along by the Bretherton flow, whose scale is much larger,
with strain~rate much less than vorticity
since $\amp^2 \ll \amp \ll \frma$ \citep[cf.][]{Kida:1981}.
That in turn makes the results
independent of detailed core structure, i.e.\ of the
function $\omega_0(\radius)$ where $\omega_0$ is the vorticity,
but dependent only on the Kelvin circulation 
$\Gamma = \int\!\!\int\!\omega_0\Sliver dxdy$.

In the case of problem~(ii), the same remote-recoil effects will be found
to occur.  In addition, because of overlap, there is a \emph{local recoil}
corresponding to advection of the vortex core by the Stokes
drift of the wavetrain.  This local contribution is given by the celebrated
Craik--Leibovich vortex force, $\omega_0$ times the Stokes drift,
equation~(\ref{eq:3dcleq-eulerian}) below,
and is directed toward negative $y$
in the case of figure~\ref{fig:probs1-2}\textit{(b)}.
Its quantum vortex counterpart is the Iordanskii force, with the Stokes drift
corresponding to the phonon current. The special limiting case where
the \ab\ effect is dominant
is also special in another way,
namely that the local contribution is the only contribution.
The remote contribution
vanishes, in that particular limit.
Generically, however,
both contributions are important,
as will be shown.

The plan of the paper is as follows.
\S\ref{sec:eqns} introduces the equations to be used,
and verifies the far-field solution (\ref{eq:disloc-wavefield}). \
\S\ref{sec:imp-psm}
recalls how the Kelvin impulse $\bimp$ of a vortex
responds to a force applied to its core, and
proves a general theorem relating the \psm\ field to the rate of change of
$\bimp$.  This is a variant of the impulse--\psm\ theorem first proved in
\citet{Buhler:2005}.  It
provides one way of seeing that the \psm\ rule holds
in all three of our problems,
independently of our explicit
calculations of wave refraction and net \psm\ flux.
The theorem does, however, depend heavily
on the smallness of $\frma$ and leaves open some challenging questions
about the wider validity
of the rule.
\S\ref{sec:breth-recoil-1} briefly revisits problem~(i), in preparation for
its extension to problem~(ii) in \S\S\ref{sec:breth-recoil-2}--7. \
In \S\ref{sec:prob3} we formulate and solve problem~(iii),
the rapidly-rotating version of problem~(i).
In that version, the waves are high-frequency deep-water surface gravity waves,
and the mean flow obeys \qg\ shallow-water dynamics in a fluid layer whose
depth $H$ is sufficiently large by comparison with $k_0^{-1}$.
The mean flow feels rotation strongly but the waves feel it only weakly.
The \oa\ wavemotion can be treated as irrotational to sufficient accuracy.
A point of interest is that the rotation produces a
tendency for the Stokes drift to be cancelled by the well-known
Eulerian-mean ``anti-Stokes flow''
\citep{Ursell:1950,
Hasselmann:1970,
Pollard:1970,
Lane:2007}.
It might be thought that the cancellation
suppresses the Bretherton flow and hence the remote recoil,
but the analysis will show otherwise. \
\S\ref{sec:conclu} offers some concluding remarks,
emphasizing
challenges for future work.

\vspace{-0.35cm}

\section{Equations used}
\label{sec:eqns}

To verify the far-field wave solution (\ref{eq:disloc-wavefield})
we need the linearized equations outside the vortex core.
The irrotational background velocity field $\bu_0$ is
$\,\bu_0(r)
\,=\,
\vspeed\sliver\radius_0\sliver\radius^{-1}\bthetavec
\,=\,
\frma\sliver\wspeed_0\sliver\radius_0\sliver\radius^{-1}\bthetavec
$\,
where $\bthetavec$ is the unit vector in the $\theta$ direction.
The equations are most succinctly written in their Bernoulli form

\vspace{-0.32cm}

\begin{equation}
\left(
   \frac{\partial}{\partial t}
   +
   \bu_0\cdott\Antisliver\bnabla
\right)
\phi'
~=~
-\wspeed^2\eta' + \chi'
~,
\label{eq:lind-phi-eqn}
\end{equation}

\vspace{-0.2cm}

\begin{equation}
\left(
   \frac{\partial}{\partial t}
   +
   \bu_0\cdott\Antisliver\bnabla
\right)
\eta'
~=~
-\nabla^2\phi'
~,
\hspace{0.6cm}
\label{eq:lind-eta-eqn}
\end{equation}
where $\phi'$ is the velocity potential for the
irrotational wavemotion, $\bu'=\bnabla\phi'$, say, while
$\eta'$ is the fractional layer-thickness or density disturbance,
in the shallow water or acoustical interpretation respectively,
and $\chi'$ is a prescribed oscillatory forcing potential.  Such forcing is a
convenient way of representing wave sources and sinks, as used in BM03 and
B14.\: In the limiting cases of problem~(ii) these sources and sinks will
recede to infinity, leaving $\chi' = 0$ for all finite $(x,y)$.  Equation
(\ref{eq:lind-eta-eqn}) is the linearized mass-conservation equation.
Eliminating $\eta'$ and
noting that 
$(\partial_t+\bu_0\cdott\Antisliver\bnabla)\Sliver\wspeed=0$,
because $\wspeed$ is a function of $\radius$ alone, we have

\vspace{-0.65cm}

\begin{equation}
\left(
   \frac{\partial}{\partial t}
   +
   \bu_0\cdott\Antisliver\bnabla
\right)^{\!\raisebox{-3pt}{\small 2}}
\phi'
~-~
\wspeed^2\nabla^2\phi'
~=~
\left(
   \frac{\partial}{\partial t}
   +
   \bu_0\cdott\Antisliver\bnabla
\right)
\chi'
~.
\label{eq:lind-wave-eqn}
\end{equation}
Now if $\,\phi'\propto \exp(\imag\sliver\Phi)$\, with \,$\Phi$\,
as in (\ref{eq:disloc-wavefield})--(\ref{eq:alpha-super-small}),
we have a local wavenumber vector

\vspace{-0.35cm}

\begin{equation}
\bk
\;=\;
\bnabla\Phi
\;=\;
k_0\bxvec - \alpha \radius^{-1}\bthetavec
+ O(\frma^2 k_0\sliver\radius_0^2\sliver\radius^{-2})
\;=\;
k_0\{
\bxvec
-
\frma\sliver\radius_0\sliver\radius^{-1}\sliver\bthetavec
+ O(\frma^2\sliver\radius_0^2\sliver\radius^{-2})
\}
\label{eq:rotated-wavenumber}
\end{equation}
where $\bxvec$ is the unit vector in the $x$ direction
and where the error term has been assumed to
have length-scale $\gtrsim k_0^{-1}$. \
We note that $\bk$ is slightly rotated away from the $x$ direction,
pointing slightly into the background flow,
as already seen in figure~\ref{fig:wavecrests-exaggerated}
where the wavecrests are
rotated away from the $y$ direction.  So

\vspace{-0.35cm}

\begin{equation}
\bnabla\phi'
\;=\;
\imag\bk\phi'
\;=\;
\imag k_0
\left\{
        \bxvec
-
        \frma\sliver\radius_0\sliver\radius^{-1}\bthetavec
+
O(\frma^2\sliver\radius_0^2\sliver\radius^{-2})
\right\}
\phi'
~,
\label{eq:grad-phiprime}
\end{equation}
and with
$\,\bu_0(r)
\,=\,
\frma\sliver\wspeed_0\sliver\radius_0\sliver\radius^{-1}\bthetavec
$
we have,
noting that $\,\bthetavec\sliver\cdott\antisliver\bxvec\,=\,\sin\theta$\,,

\vspace{-0.25cm}

\begin{equation}
\left(
   \frac{\partial}{\partial t}
   +
   \bu_0\cdott\Antisliver\bnabla
\right)
\phi'
\;=\;
\imag\sliver k_0\sliver\wspeed_0
\left\{
- 1
+
\frma\sliver\radius_0\sliver\radius^{-1}\sin\theta
+
O(\frma^2\sliver\radius_0^2\sliver\radius^{-2})
\right\}
\phi'
\label{eq:d-phiprime}
\end{equation}
and

\vspace{-0.65cm}

\begin{equation}
\left(
   \frac{\partial}{\partial t}
   +
   \bu_0\cdott\Antisliver\bnabla
\right)^{\!\raisebox{-3pt}{\small 2}}
\phi'
\;=\;
-k_0^2\wspeed_0^2
\left\{
  1
  - 2\frma\sliver\radius_0\sliver\radius^{-1}\sin\theta
  + O(\frma^2\sliver\radius_0^2\sliver\radius^{-2})
\right\}
\phi'
~,
\label{eq:d2-phiprime}
\end{equation}
which equals
$\,\wspeed_0^2\nabla^2\sliver\phi'\,=\,-|\bk|^2\wspeed_0^2\Sliver\phi'$\,
to the same accuracy and therefore
satisfies (\ref{eq:lind-wave-eqn}) with $\,\chi'\,=\,0$,\,
to that accuracy.
The next order $\,O(\frma^2\sliver\radius_0^2\sliver\radius^{-2})$\,
fails to satisfy (\ref{eq:lind-wave-eqn}), because a contribution
$-2\sliver k_0^2\wspeed_0^2\sliver
\frma^2\sliver\radius_0^2\sliver\radius^{-2}\phi'$
on the right of (\ref{eq:d2-phiprime})
disagrees with a contribution
$-k_0^2\wspeed_0^2
\frma^2\sliver\radius_0^2\sliver\radius^{-2}\phi'$
to
$\,\wspeed_0^2\nabla^2\sliver\phi'\,$,
  with no coefficient 2.
At higher orders there are contributions from
$\bnabla(\sin\theta)$
that also disagree.
We note in passing that, by contrast,
$\exp(\imag\sliver\Phi)$\, with no error term
is an exact, and not merely a far-field asymptotic, solution to the
Schr\"odinger equation
of the original \ab\ problem (details in Appendix~A).
The Schr\"odinger equation (\ref{eq:schro-eqn})
differs from (\ref{eq:lind-wave-eqn}) except in
the limit $\frma \rightarrow 0$
\citep[e.g.][]{Stone:2000a}.

The expression (\ref{eq:rotated-wavenumber}) for $\bk$
is consistent with ray (JWKB) theory,
as can be checked from BM03 (4.12) or B14 (14.5).
Also of interest is the direction of the
absolute group velocity

\vspace{-0.65cm}

\begin{equation}
\bCabs = \frac{\wspeed\sliver\bk}{|\bk|} + \bu_0(\radius)
 = \frac{\wspeed_0\sliver\bk}{|\bk|} + \bu_0(\radius)
+ O(\frma^2\wspeed_0\sliver\radius_0^2\sliver\radius^{-2})
~.
\label{eq:abs-gp-vel}
\end{equation}
Correct to $O(\frma\wspeed_0\sliver\radius_0\sliver\radius^{-1})$,
$\bCabs$ is parallel to $\bxvec$,
as can be checked by taking the $y$ components of
(\ref{eq:rotated-wavenumber})
and of
$\,\bu_0(r)
\,=\,
\frma\sliver\wspeed_0\sliver\radius_0\sliver\radius^{-1}\bthetavec
$.
Propagation due to the $y$ component of $\bk$ cancels
advection due to the $y$ component of
$\,\bu_0$. \
The cancellation
follows also from the irrotationality of the background flow outside the
vortex core, in virtue of
the curl-curvature formula of ray theory, B14 p.\;86.

We avoided relying on ray theory here,
when verifying (\ref{eq:disloc-wavefield}),
because integrating the ray equations
over large distances might, conceivably,
accumulate significant errors in $\Phi$,
giving incorrect values for the \abpj,
whereas the error $O(\frma^2\radius_0^2/\radius^2)$
in (\ref{eq:disloc-wavefield})
is small enough to rule out any such accumulation.
A convenient corollary is that (\ref{eq:disloc-wavefield})
can be used in problem~(i) as well as in problem~(ii), because when
ray theory is valid it is permissible, correct to \oeps,
to replace the constant amplitude
$\envel$ by a $y$-dependent amplitude envelope that
restricts the wavetrain appropriately, as sketched in
figure~\ref{fig:probs1-2}\textit{(a)}, again using the \oeps\ property
$\bCabs \:||\: \bxvec$ just noted
(as contrasted with the $O(\frma^2)$ bending of ray paths in
figure~\ref{fig:bretherton}).

Ray theory will, on the other hand, be sufficient for our treatment of
problem~(iii), in which the \ab\ phase jump has no role.
Details are postponed until \S\ref{sec:prob3}.

For the mean flow,
a natural and efficient
framework for solving problems~(i)--(iii)
is that of generalized Lagrangian-mean (GLM) theory,
as laid out for instance in B14.
However,
except where stated otherwise
the reader unfamiliar with GLM theory can read the equations as
involving, to sufficient accuracy, only the Eulerian-mean velocity
$\bubar$ and the
Stokes drift $\bubars$, or phonon current per unit mass.
Whenever the \oa\ wavemotions are irrotational and
describable by ray theory,
the exact GLM \psm\ $\bpmom$ per unit mass can be
replaced by  $\bubars$
and the exact Lagrangian-mean velocity $\bubarl$
by  $\bubar + \bubars$, with error \oaaee;
see e.g.\ B14, equations~(10.15) and (10.17).
Then the combination  $\bubarl - \bpmom$,  which occurs frequently
in the exact theory,
can be read simply as~$\bubar$, and the
exact mean vorticity $\bomegatilde$ defined by

\vspace{-0.45cm}

\begin{equation}
\bomegatilde
\;=\;
\bnabla\cross(\bubarl - \bpmom)
\label{eq:exact-mean-vorticity}
\end{equation}
can be read simply as~$\bomegabar$, the Eulerian-mean vorticity.
(The quantity $\bomegatilde$ is the simplest exact measure of mean
vorticity.  It arises from frozen-field distortions of the
three-dimensional vorticity field by
the wave-induced displacement field.) \
The relation $\bpmom = \bubars$ is always
valid sufficiently far from the vortex core, in all three problems,
where ray theory is always valid.
In problems~(i) and~(ii) we need only the vertical or $z$ component of
(\ref{eq:exact-mean-vorticity}).

The power and economy of the GLM formalism comes from
Kelvin's circulation theorem and its consequence,
e.g.\ B14 \S10.2.9,
that $\bubarl$ exactly advects mean vorticities $\bomegatilde$,
or $\bomegatilde + \bfcoriolis$ in problem~(iii),
with $\bfcoriolis$ the vector Coriolis parameter.
This will prove useful throughout our analyses.  In
problem~(iii) it expresses Ursell's insight into
the anti-Stokes flow,
in a succinct and natural way to be pointed out in \S\ref{sec:prob3}.
The advection property is neatly summarized by the
exact three-dimensional form of the nondissipative
equation for $\bomegatilde$; see e.g.\
B14, equations~(10.99) and (10.153). \
It is

\vspace{-0.25cm}

\begin{equation}
\frac{\partial\bomegatilde}{\partial t}
~-~
\bnabla\cross\{\bubarl\cross(\bfcoriolis + \sliver\bomegatilde)\}
~=~
0
\label{eq:3d-exact-mean-vort}
\end{equation}
or alternatively

\vspace{-0.35cm}

\begin{equation}
\frac{\Dbarl\bomegatilde}{Dt}
~+~
(\bfcoriolis + \sliver\bomegatilde)\Sliver\bnabla\cdott\bubarl
~=~
(\bfcoriolis + \sliver\bomegatilde)\sliver\cdott\Antisliver\bnabla\bubarl
,
\label{eq:3d-exact-mean-vort-cdot-form}
\end{equation}
where
$\Dbarl/Dt = \partial/\partial t + \bubarl\cdott\bnabla$. \
If we now set $\bfcoriolis=0$
and apply the foregoing recipe to (\ref{eq:3d-exact-mean-vort}),
replacing
$\bomegatilde$ by~$\bomegabar$
and
$\bubarl$ by $\bubar + \bubars$,
then we get the approximate version of (\ref{eq:3d-exact-mean-vort})
known as the \cleq:

\vspace{-0.25cm}

\begin{equation}
\frac{\partial\bomegabar}{\partial t}
-
\bnabla\cross(\bubar\cross\bomegabar)
~=~
\bnabla\cross(\bubars\cross\bomegabar)
~.
\label{eq:3dcleq-eulerian}
\end{equation}

\smallskip
\noindent
The \rhs\ of (\ref{eq:3dcleq-eulerian})
is the curl of the \clvf\ $\bubars\cross\bomegabar$,
which as mentioned earlier makes a local contribution 
$\bubars\cross\bomega_0$
to the effective force on the vortex
in problem~(ii), where
$\bomega_0 = \bnabla\cross\bu_0$.
Equation (\ref{eq:3dcleq-eulerian})
was originally derived by
\citet{Craik:1976}, via a much longer route,
to study another problem -- the dynamics of
Langmuir vortices --
assuming incompressible flow $\bnabla\cdott\bu=0$
and steady wave~fields, and
under the strong parameter restriction
$\amp^2 \:\sim\: \frma \:\ll\: 1$, 
i.e.\ that all mean flows, whether wave-induced or
pre-existing, have order of magnitude \oaa.
The route via GLM just recalled, which is not only much shorter but also
has wider validity,
was first pointed out by \citet{Leibovich:1980}.

To complete the mean-flow equations we need a mass-conservation equation.
As usual in GLM, we define a mean two-dimensional density or layer depth
$\htilde$ such that the areal mass element $\propto\htilde\Sliver dxdy$
exactly; see equations~(10.42)--(10.47) of B14.
Then mass conservation is expressed by

\vspace{-0.65cm}

\begin{equation}
\frac{\partial\htilde}{\partial t}
~+~
\bnablah\cdott(\htilde\Sliver\bubarlh)
~=~
0
~,
\label{eq:mean-mass-conservation}
\end{equation}
where suffix {\small H} denotes horizontal projection,
on to the $xy$ plane, superfluous in the
two-dimensional acoustical setting but needed in the
shallow-water setting and in problem~(iii).
In all three problems, however, we shall find that
(\ref{eq:mean-mass-conservation}) can be simplified to

\vspace{-0.35cm}

\begin{equation}
\bnablah\cdott\bubarlh
~=~
0
\label{eq:mean-mass-nondiv}
\end{equation}
if we are willing to work to the lowest significant
accuracy -- the lowest
that captures the remote-recoil effects to be discussed.
This will keep the analysis extraordinarily simple
yet able to illustrate the main points.
Equation~(\ref{eq:mean-mass-nondiv}) will be justified shortly
for problems~(i) and~(ii) and in \S\ref{sec:prob3}
for problem~(iii), with error estimates.
We can then define a streamfunction, $\psitilde$ say, such that the
horizontal velocity components can be written as

\vspace{-0.65cm}

\begin{equation}
\ubarlh
~=~
-\frac{\partial\psitilde}{\partial y}
\qquad \mbox{and} \qquad
\vbarlh
~=~
\frac{\partial\psitilde}{\partial x}
~.
\label{eq:streamfunctiondef}
\end{equation}
The
Bretherton flow
has streamfunction

\vspace{-0.39cm}

\begin{equation}
\psitildeb
~=~
\psitilde
~-~
\psitilde_0
\label{eq:bstreamfunctiondef}
\end{equation}
where $\psitilde_0$ is the streamfunction for
the nondivergent velocity field
$\bu_0$ of the vortex flow.

To compute the Bretherton flows in problems (i) and (ii)
to sufficient accuracy
(see \S4),
we need only (\ref{eq:streamfunctiondef})--(\ref{eq:bstreamfunctiondef}) and
the vertical component of (\ref{eq:exact-mean-vorticity}).  We have
$\bomegatilde = 
\bnabla\cross(\bubarl - \bpmom) = \bnabla\cross\bu_0
=\bomega_0$\sliver, \
expressing irrotationality outside the vortex core.
The vertical component of~$\bnabla\cross(\bubarl - \bu_0)$ is just
$\nablahsq\psitildeb$. \
We therefore have

\vspace{-0.35cm}

\begin{equation}
\nablahsq\psitildeb
~=~
\bzvec\sliver\cdott\Antisliver\bnabla\cross\bpmom
\label{eq:for-probs-1-2}
\end{equation}
where $\bzvec$ is the vertical unit vector.
The \rhs\ of (\ref{eq:for-probs-1-2}) is known
as soon as we know the wave \psm\ field $\bpmom$, which
as mentioned earlier can be identified with
the $\bubars$ field whenever the
wavemotion is irrotational and ray theory applies.

For problem~(iii) it will be shown in \S\ref{sec:prob3}
that we need only two modifications.  First, the vorticity
$\nablahsq\psitildeb$ must be replaced in the standard way by
$(\nablahsq - \LD^{-2})\psitildeb$,
the \qg\ potential vorticity (PV),
of the Bretherton flow,
where $\LD$ is the Rossby deformation length-scale
$\LD=\fcoriolis^{-1}(gH)^{1/2}$
where $g$ is gravity and $\fcoriolis = |\bfcoriolis|$.
Second, we must replace $\bpmom$, which
for deep-water surface gravity waves is strongly $z$-dependent,
by its vertical average $\langle\bpmom\rangle$. \
So in place of (\ref{eq:for-probs-1-2}) we have simply

\vspace{-0.35cm}

\begin{equation}
(\nablahsq - \LD^{-2})\psitildeb
~=~
\bzvec\sliver\cdott\Antisliver\bnabla\cross\langle\bpmom\rangle
~.
\label{eq:for-prob-3}
\end{equation}
The derivation of (\ref{eq:for-prob-3}) involves some delicate
arguments about the asymptotics and will be postponed until
\S\ref{sec:prob3} and Appendix~\ref{sec:appc}.
The elliptic operators in
(\ref{eq:for-probs-1-2}) and (\ref{eq:for-prob-3})
illustrate, by implication, a generic property of Bretherton flows,
that they can extend well outside the wavetrain where $\bpmom\ne0$.
That is one way of seeing
the generic nature of remote recoil.

We now justify replacing (\ref{eq:mean-mass-conservation}) by
(\ref{eq:mean-mass-nondiv}) for problems~(i) and~(ii).
Among the errors thus incurred,
the largest
is \oaae.
It arises in problem~(i), from the variation of wave amplitude $A$
across the wavetrain and illustrating, incidentally, a need to avoid
textbook arguments for the
near-incompressibility of low-Mach-number or low-Froude-number flows.
Those arguments do not take into account the kinds of spatial heterogeneity
that are possible here,
especially in problem~(i).

For our asymptotic regime we need to let
$\amp\rightarrow 0$ and $\frma\rightarrow 0$ keeping $\amp\ll\frma$,
for a given geometry of the background flow and incident wave~field.
Where the vortex flow $\bu_0(\radius)$ crosses the wavetrain,
in problem~(i),
it encounters $\htilde$~values that are reduced by \oaaeo\
because of the Brillouin radiation stress in the wavetrain.
And since $\bnablah\cdott\bu_0=0$, the resulting contribution
to $\bnablah\cdott(\htilde\sliver\bubarlh)$
in (\ref{eq:mean-mass-conservation}),
which is neglected in going to
(\ref{eq:mean-mass-nondiv}),
is just
$\bu_0\Sliver\cdott\bnablah\htilde$
to leading order, with magnitude \oaae\
since $\bu_0=$ \oeps.
(This contribution is significant, however,
at the greater accuracy required in the case of
figure~\ref{fig:bretherton},
as can be seen from (5.1) of BM03,
even though it will not be required in the
present analyses.)

The \oaaeo\ local reduction in $\htilde$ within the wavetrain
(set-down, in the shallow-water setting)
is necessary to ensure that
the \oaa\ sideways mean fluid acceleration vanishes.
The sideways gradients of Brillouin radiation stress and
\oaa\ mean pressure must cancel.  Only
the isotropic part of the Brillouin radiation stress
is involved,
the so-called
``hard-spring'' contribution,
unrelated to \psm\ fluxes, and
equal to $\partial\ln c/\partial\ln h$ times wave-energy per unit area,
where $h$ is layer depth or two-dimensional mass density
\citep[e.g.][B14 \S10.5.1]{Brillouin:1936}.

\vspace{-0.3cm}

\section{Impulse and \psm}
\label{sec:imp-psm}

The impulse--\psm\ theorem applies to problems~(i)--(iii) 
as well as to a more general set of problems involving multiple vortices
and more complicated wave~fields, with arbitrary wave source and sink regions.
The theorem provides an elegant way of showing
that the \psm\ rule is satisfied in all these problems.
There is, however, a severe limitation.
The theorem relies crucially on horizontal nondivergence,
(\ref{eq:mean-mass-nondiv}).
  So it applies only at the lowest significant order of accuracy.
There is a challenge here since the limitation puts the case of
figure~\ref{fig:bretherton} outside the scope of the theorem.
As just pointed out, (5.1) of BM03 shows that
(\ref{eq:mean-mass-nondiv}) is not accurate enough in that case; in fact
(\ref{eq:mean-mass-nondiv}) must then be replaced by the anelastic equation

\vspace{-0.35cm}

\begin{equation}
\bnablah\cdott(\sliver\htilde\Sliver\bubarlh)
~=~
0
~.
\label{eq:mean-mass-anelastic}
\end{equation}
Yet
BM03's analysis shows that the \psm\ rule still holds, a point to which we
will return.

Before proceeding, we revisit the \te\ in which the waves are removed and an
artificial external force field $\bF$ exerted on the vortex core
$\omega_0(\radius)$, producing a rate of change of Kelvin impulse.
In order to make the vortex translate bodily
without change of shape, with velocity $\butr$, say,
we need
$\bF = -\sliver\omega_0\Sliver\bzvec\sliver\cross\butr$.
(The curl of this force field
is just that required to move the vortex core
through the fluid at velocity $\butr$,
while the divergence sets up the dipolar pressure field required
to produce
the corresponding
changes outside
the core,
where the velocity field is irrotational.) \
Being transverse to the vortex motion, the resultant force 
$\bR$ has the character of a Magnus force,

\vspace{-0.35cm}

\begin{equation}
\bR 
\;=\;
\int\!\!\int\!\bF\Sliver dxdy
\;=\; -\sliver
\bzvec\sliver\cross\butr\Antisliver\int\!\!\int\!\omega_0\Sliver dxdy
\;=\; - \sliver\Gamma\sliver\bzvec\sliver\cross\butr 
~.
\label{eq:magnus-force}
\end{equation}
The Kelvin impulse is defined for our
two-dimensional shallow water or acoustical domain as

\vspace{-0.67cm}

\begin{equation}
\bimp
~=\;
\int\!\!\int
(y,\,-x)\Sliver Q
\Sliver dxdy
~=\;
\int\!\!\int -
\bzvec\sliver\cross\bx\Sliver\Sliver Q
\Sliver dxdy
\label{eq:impulse-def}
\end{equation}
\citep[e.g.][equation~(7.3.7)]{Batchelor:1967}
where $\bx=(x,y)$
and where $Q$ is the vorticity,
$Q = \omega_0$ in this case.
When the vortex translates in response to $\bF$,
the rate of change of $\bimp$ is just $\bR$; cf.\ (\ref{eq:impulse-evol}) below.
In the corresponding \te\ for problem~(iii),
the same statements hold if $Q$ is redefined as the \qg\ PV.

The impulse--\psm\ theorem makes the following assumptions,
in addition to (\ref{eq:mean-mass-nondiv}) and its consequences
(\ref{eq:streamfunctiondef})--(\ref{eq:for-prob-3}).
The wave~field together with its sources and sinks is taken to have finite
extent, prior to taking any infinite-wavetrain limits
that might be of interest,
while the domain of integration is taken infinite so as to enclose within
it the vortex core, or cores, as well as all the waves and their source and
sink regions.
It is assumed that the \psm\ field satisfies a two-dimensional equation
of the form (see Appendix~\ref{sec:appd})

\vspace{-0.35cm}

\begin{equation}
\frac{\partial\bpmom}{\partial t}
~+~
\bnablah\cdott\bfpmom
~=\;
-\,
(\bnablah\bubarl)\sliver\cdott\bpmom
~+~
\pmb{\cal F}
~,
\label{eq:psm-law}
\end{equation}
with vertical averaging understood in problem~(iii).
The first term on the right comes
from wave refraction and scattering
by the mean flow,
and
$\pmb{\cal F}$ is
the rate of generation or absorption
of \psm\ in the wave source and sink regions, per unit area.
In the refraction term,
$\bpmom$ contracts with $\bubarl$ and not with $\bnablah$\!. \
On the left,
the precise form of the \psm\ flux tensor $\bfpmom$
is immaterial but we note for later reference
that, wherever
ray theory holds, we shall have, in Cartesian tensor notation,
with $i$ and $j$ running from 1 to 2,
the standard group-velocity property

\vspace{-0.35cm}

\begin{equation}
\fpmom_{ij}
~=~
\pmom_i\,\Cabs_j
\label{eq:pmom-flux-gp-vel}
\end{equation}
where $\bCabs$ is the absolute group velocity,
$\bu_0$ plus the intrinsic group velocity, $c\bk/|\bk|$ in problems~(i) and (ii)
as in (\ref{eq:abs-gp-vel}),
and
$c\bk/2|\bk|$ in problem~(iii).
The divergence operator contracts with $\bCabs$
so that, in Cartesians, the $i$th component of $\bnablah\cdott\bfpmom$
is \,$\fpmom_{ij,j}$.
The
theorem states that

\vspace{-0.65cm}

\begin{equation}
\framebox{$\displaystyle
\frac{d}{dt}
\left(
   \bimp ~+~ \bpmomint
\right)
~=\;
\int\!\!\int
  \pmb{\cal F}\;
dxdy$}
\label{eq:imp-psm-theorem}
\end{equation}
where $\bpmomint = \int\!\!\int\!\bpmom\Sliver dxdy$, the total \psm,
again with vertical averaging understood in problem~(iii).

The proof begins by noting that
$Q\Sliver dxdy$ is materially invariant, so that

\vspace{-0.35cm}

\begin{equation}
\frac{d\sliver\bimp}{dt}
~=~
\!\!\int\!\!\int
\frac{\Dbarlh(y,\,-x)}{Dt}
\Sliver Q
\Sliver dxdy
~=~
\!\!\int\!\!\int
(\vbarlh,\,-\ubarlh)\Sliver Q
\Sliver dxdy
~.
\label{eq:impulse-evol}
\end{equation}
From here on,
with everything in two dimensions $(x,y)$, we
drop the suffixes {\small H} so that, for instance,
$\bnablah$ will be denoted by $\bnabla$. \ Then,
recalling
(\ref{eq:exact-mean-vorticity}) and
(\ref{eq:streamfunctiondef})--(\ref{eq:for-prob-3}), we have

\vspace{-0.3cm}

\begin{equation}
\frac{d\sliver\bimp}{dt}
~=~
\!\!\int\!\!\int
Q
\Sliver\bnabla\psitilde
\, dxdy
~=~
\!\!\int\!\!\int
\left\{
(\nablasq - \LD^{-2})\psitilde
-\, \bzvec\sliver\cdott\Antisliver\bnabla\cross\bpmom
\right\}
\bnabla\psitilde
\, dxdy
~,
\label{eq:impulse-evol-bis}
\end{equation}
with $\LD$ finite in problem~(iii)
but infinite in problems~(i)--(ii).
Now
$\nablasq\psitilde\sliver\bnabla\psitilde$
contributes nothing
because, in Cartesians,
its $i$th component is
$\psitilde_{,jj}\sliver\psitilde_{,i}
=
(\psitilde_{,j}\sliver\psitilde_{,i})_{,j}
-\half
(\psitilde_{,j}\sliver\psitilde_{,j})_{,i}
$\Sliver,
which integrates to zero.
The integrated terms at infinity vanish because, if we consider
a domain of integration having radius
$\radius \rightarrow \infty$,
the integrated terms
have integrands
\,$O(\radius^{-2})$\, in problems (i) and (ii) and
\,$O(\exp(-2\radius/\LD))$\, in problem~(iii), from
the vortex-only contributions.
The Bretherton flows, being dipolar
because of the $\bnabla$ on the \rhs s of
(\ref{eq:for-probs-1-2}) and (\ref{eq:for-prob-3}),
decay at the same rate or faster as
$\radius \rightarrow \infty$.
In problem~(iii) we have the additional contribution
$-\LD^{-2}\psitilde
\sliver\bnabla\psitilde \;\propto\; \half\bnabla(\psitilde^2)$,
which also integrates to zero because $\psitilde^2=O(\exp(-2\radius/\LD))$.
Therefore
(\ref{eq:impulse-evol-bis}) reduces,
in all three problems, to

\vspace{-0.35cm}

\begin{equation}
\frac{d\sliver\bimp}{dt}
~=\,
-\antisliver
\int\!\!\int
   (\bzvec\sliver\cdott\Antisliver\bnabla\cross\bpmom)
   \Sliver
   \bnabla\psitilde\,
dxdy
~.
\label{eq:impulse-evol2}
\end{equation}
Upon exchanging the dot with the cross and then
integrating by parts,
using the finite extent of the wave~field,
we see that the \rhs\ is equal to

\vspace{-0.3cm}

\begin{equation}
-\antisliver
\!\!\int\!\!\int
   \{(\bzvec\Sliver\cross\Antisliver\bnabla)\sliver\cdott\bpmom\}
   \bnabla\psitilde\,
dxdy
~=~
\!\!\int\!\!\int
   \{\bpmom\sliver\cdott(\bzvec\Sliver\cross\Antisliver\bnabla)\}
   \bnabla\psitilde\,
dxdy
~=~
\!\!\int\!\!\int
   (\bnabla\bubarl)\cdott\bpmom\,
dxdy
\label{eq:last-int-by-parts}
\end{equation}
since
$\bzvec\Sliver\cross\Antisliver\bnabla$ commutes with
$\bnabla$, and
$\bzvec\Sliver\cross\Antisliver\bnabla\psitilde = \bubarl$ by
(\ref{eq:streamfunctiondef}), so that the integrand on the right
is minus the refraction term in (\ref{eq:psm-law}).
On eliminating
that term between (\ref{eq:psm-law}) and
(\ref{eq:last-int-by-parts}),
and noting that $\bnabla\cdott\bfpmom$ integrates to zero,
again because of the finite extent of the wave~field,
we
arrive at
(\ref{eq:imp-psm-theorem}).

The proof has no dependence on whether or not ray theory applies.
It depends only on (\ref{eq:mean-mass-nondiv})--(\ref{eq:for-prob-3})
and on the form of
(\ref{eq:psm-law}),
not on any particular formulae for $\bpmom$, $\bfpmom$ and
$\pmb{\cal F}$.
The group-velocity property
(\ref{eq:pmom-flux-gp-vel})
will, however, be useful when considering \psm\ budgets in detail,
for problems~(i)--(iii),
because along with ray theory it always holds far from the vortex core,
together with the approximation $\bpmom = \bubars$.

The case of figure~\ref{fig:bretherton} prompts the question,
can we replace (\ref{eq:mean-mass-nondiv}) by
(\ref{eq:mean-mass-anelastic}) and still prove the theorem?
After considerable effort, the author has been
forced to the conclusion that we cannot.  At higher
orders in $\frma$ there is
an incompatibility
between the per-unit-mass basis of vorticity --
being the curl of velocity rather than of mass transport --
and the per-unit-volume, or per-unit-area, basis of
conservation relations, and
their refractive extensions such as (\ref{eq:psm-law}),
in which $\bpmom$ would need to be replaced by $\htilde\sliver\bpmom$
in order to attain enough accuracy to be compatible with
(\ref{eq:mean-mass-anelastic}) (again see Appendix~\ref{sec:appd}).
But one cannot simply insert a factor $\htilde$ into the integrand of
(\ref{eq:impulse-def}),
because
$\htilde\sliver Q\Sliver dxdy$ is not materially invariant
and the subsequent steps from (\ref{eq:impulse-evol}) onward
are invalidated. 

It seems likely that this limitation is
not a limitation of the \psm\ rule as such,
but only a limitation of the Kelvin impulse concept.
As is well known, the ability to replace momentum budgets by impulse
budgets depends on incompressibility.
Incompressibility is needed in order to
banish to infinity the large-scale
\oaa\ pressure-field adjustments in transient situa\-tions, including
transient versions of \te s
like that associated with the Magnus relation (\ref{eq:magnus-force}).

Indeed, there is a variant of the case of
figure~\ref{fig:bretherton} in which the \psm\ rule holds to all
orders in $\frma$, as was shown in BM03 \S5.2.  However, that result depended
on keeping the mean flow exactly steady -- over an infinite domain --
by applying an artificial ``holding force'' to the vortex core as described in
\S\ref{sec:intro}, and then taking account of the full momentum budget
in the far field correct to \oaa.  Conditions in the far field are greatly
simplified by assuming exact steadiness everywhere.  The result depends on
the generic relation between the fluxes of momentum and \psm.  As shown in
GLM theory they differ only by an isotropic term (see Appendix~\ref{sec:appd}),
which can be
balanced by changes in mean pressure.
Whether that can lead to further generalization remains to be explored.

\vspace{-0.45cm}

\section{Bretherton flows and recoil forces for problem~(i)}
\label{sec:breth-recoil-1}

In this section we review BM03's leading-order results
on recoil forces in problem~(i),
as a preliminary to the subsequent analyses of problem~(ii).
We confine attention to recoil forces
computed from Bretherton flows, omitting BM03's
ray-theoretic calculations
of the \oaae\ net \psm\ flux.
The impulse--\psm\ theorem tells us that
such calculations
must give the same results, at this lowest order,
as indeed they were found to do.
However, as pointed out in BM03,
the Bretherton flows provide the simplest route to the results.
Just as $\bR$
in (\ref{eq:magnus-force})
can be computed correct to \oaae\ from $\butr$ correct to \oaaeo,
because $\Gamma$ is an \oeps\ quantity,
a wave-induced recoil can be computed correct to \oaae\ from a
Bretherton flow correct
to \oaaeo, that is, from a Bretherton flow computed for
an \emph{unrefracted} wavetrain.

To take advantage of this simplification, in problem~(i), we need
to consider a
wavetrain of finite length.  That is because of the curl-curvature formula
mentioned below (\ref{eq:abs-gp-vel}), with its implication that
the absolute group velocity $\bCabs$
remains parallel to $\bxvec$ when wave refraction is computed correct to \oeps.
It remains parallel
despite the \oeps\ refraction effects
illustrated in figure~\ref{fig:wavecrests-exaggerated};
recall the cancellation noted below (\ref{eq:abs-gp-vel}).
The \oaae\ net \psm\ flux and recoil force,
which depend entirely on \oeps\ wavecrest rotations of the kind
illustrated in figure~\ref{fig:wavecrests-exaggerated},
therefore vanish
when the wave source and sink are allowed to recede to
infinity.

\begin{figure} 
\centering
\vspace{0.1cm}
\hspace{0.1cm}
\scalebox{0.95}[1]{\includegraphics[width=9cm, viewport=0 300 800
700,clip]{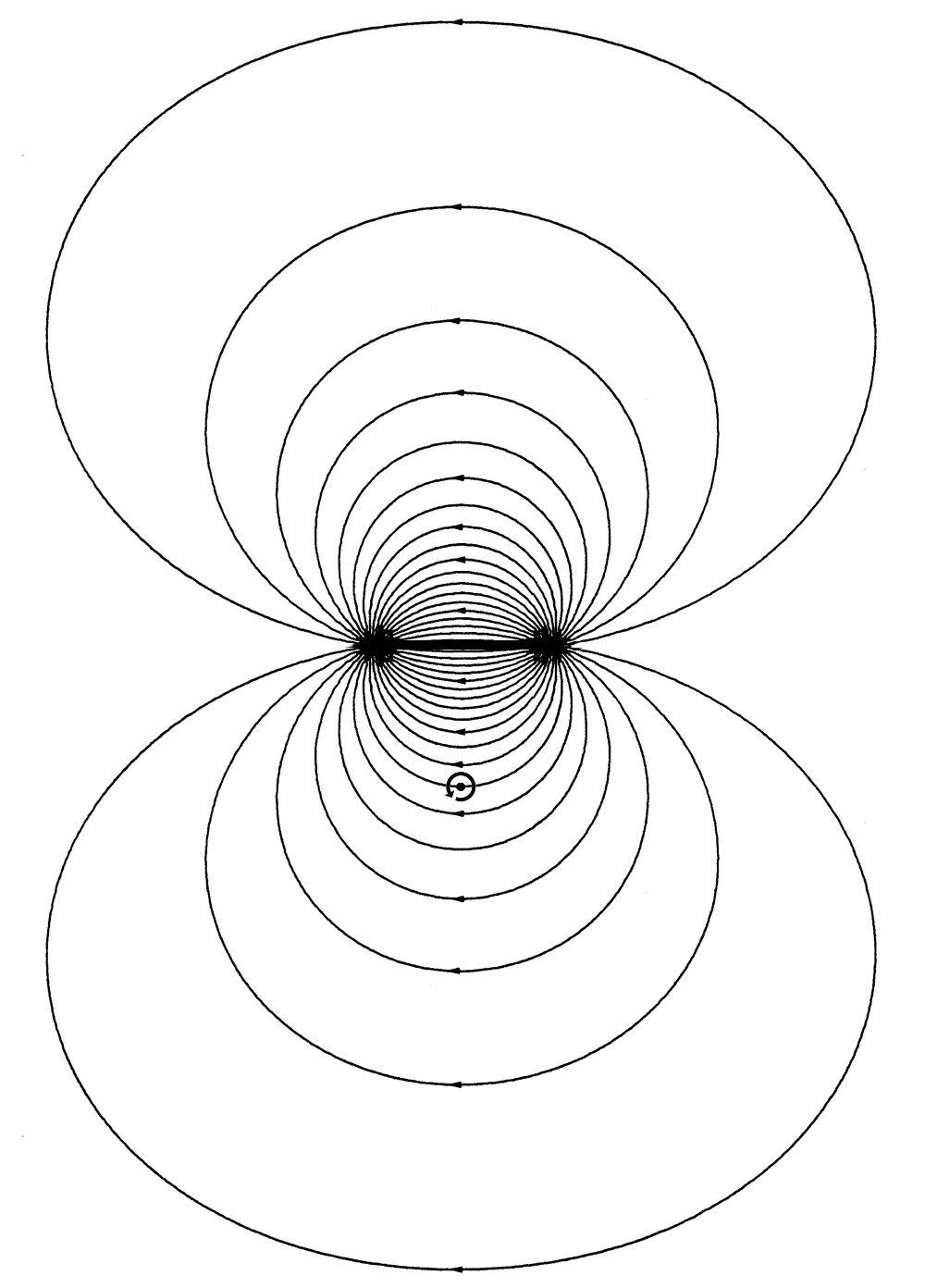}}
\caption{Schematic of Bretherton-flow streamlines in problem~(i), as analysed
in BM03 correct to lowest order \oaaeo\ for the finite wavetrain whose ray
path is shown by the heavy straight line.  At this order the Stokes
drift is nondivergent except within the wave source and sink regions.
The waves propagate from a source on the left to a sink on the right.
}
\vspace{-0.35cm}
\label{fig:bretherton-finite}
\end{figure}

So to obtain the simplest, leading-order version of
problem (i),
following BM03,
we let the unrefracted wavetrain extend between
wave source and sink regions centred at finite locations
$(x,\,y) = (-X, Y)$ and $(+X, Y)$, say,
as sketched in figure~\ref{fig:bretherton-finite} (heavy straight line).
For consistency with BM03's use of ray theory we take
$X,Y \Antisliver\gg \width$\,,
\,and $\width \Antisliver\gg k_0^{-1}$, where $\width$ is a width-scale for the wavetrain.
In order to generate and absorb an approximately monochromatic wavetrain,
the wave source and sink
are prescribed, just as in BM03, by taking
the oscillatory forcing potential $\chi'$ to be
$\exp(\imag \bk\Sliver\cdott\bx -\imag kct)$ times
a slowly-varying forcing envelope whose
length-scale is at least of the
same order of magnitude as $\width$,
and where the real part is understood.
However, the forcing envelope scale is kept $\ll X, Y$, allowing us to
think of the source and sink regions as approximately localized.
It is convenient to take $\bk\Sliver\cdott\bx = k_0x$ in the source
along with $kct = k_0ct$, with constant $k_0$.

The \oaaeo\ Bretherton flow
for the unrefracted wavetrain satisfies
(\ref{eq:for-probs-1-2}), with evanescence at infinity.
It consists of the Stokes drift $\bubars$ straight along the wavetrain,
parallel to $\bxvec$, together with irrotational return flows
symmetrically on both sides,
as sketched in figure~\ref{fig:bretherton-finite}.
All the vorticity of this Bretherton flow comes from
$\bzvec\sliver\cdott\Antisliver\bnabla\cross\bpmom$
on the \rhs\ of (\ref{eq:for-probs-1-2}).
Within the wavetrain but outside the wave source and sink,
$\bzvec\sliver\cdott\Antisliver\bnabla\cross\bpmom$
is simply minus the horizontal shear of the Stokes drift.

In virtue of the incompressibility expressed by
(\ref{eq:mean-mass-nondiv})--(\ref{eq:bstreamfunctiondef}),
the irrotational return flow outside the wavetrain, 
which advects the vortex core, is
the same as if it were induced by a two-dimensional
mass sink at the left end of the wavetrain and a mass source at the
right end whose strengths
are equal to the mass flow in the
Stokes drift within the wavetrain.
Omitting factors $\rho$, where $\rho$ is fluid mass density,
we define the source and sink strengths
$\pm S$ as the volume fluxes in a layer of unit depth;
thus

\vspace{-0.35cm}

\begin{equation}
S = \int \us(y)\sliver dy
  = \int \pmom_1(y)\sliver dy
~,
\label{eq:mass-flow-prob-1}
\end{equation}
with the integral taken across the wavetrain.
The \psm\ within the
wavetrain has been written correct to \oaaeo\ as
$\pmom_1(y)\bxvec$.
Using $\width \Antisliver\ll X$ and $\width \Antisliver\ll Y$, again following BM03,
we can approximate the
mass-source flow in problem~(i) as radially outward from
$(x,\,y) = (X, Y)$
at speed $S/[2\pi\{(x-X)^2+(y-Y)^2\}^{1/2}]$,
and similarly the mass-sink flow as radially inward toward
$(x,\,y) = (-X, Y)$.
When these flows are added vectorially we obtain the flow pattern
sketched in figure~\ref{fig:bretherton-finite};
and the net velocity advecting the vortex core at
$(x,\,y) = (0, 0)$ is

\vspace{-0.75cm}

\begin{equation}
\bubarl(0,0)
~=~
\frac{S}{\pi}
\sliver
\frac{X}{X^2+Y^2}
\sliver(-\bxvec)
\label{eq:core-advec-prob-1}
\end{equation}
correct to \oaaeo.  Because of the Magnus relation this
corresponds to a resultant recoil
force $\bR = -\Gamma\sliver\bzvec\sliver\cross\bubarl(0,0)$, i.e.\

\vspace{-0.55cm}

\begin{equation}
\bR
~=~
\frac{\Gamma S}{\pi}
\sliver
\frac{X}{X^2+Y^2}
\sliver(+\byvec)
\label{eq:force-prob-1}
\end{equation}
correct to \oaae,
where $\byvec$ the unit vector in the $y$ direction.

Notice that 
$\bubarl(0,0)$ and $\bR$ tend toward zero in the formal limit
$X\rightarrow\infty$, cross-checking what was deduced from the
direction of $\bCabs$ and consequent 
vanishing of the \oaae\ net \psm\ flux and recoil force
in that limit.
As the wavetrain gets longer, the irrotational return
part of the Bretherton flow becomes
increasingly spread out in the $y$ direction, diluting its effect
at $(x,\,y) = (0, 0)$.

\vspace{-0.35cm}

\section{Bretherton flows and recoil forces for problem~(ii)}
\label{sec:breth-recoil-2}

The dilution effect just pointed out
is the easiest way of seeing
the \nil\
in problem~(ii).
The same dilution effect occurs
for any wavetrain whose width $\width$ is held fixed while its length
$\length=2\xextent\rightarrow\infty$.
In that formal limit
the magnitude of the return flow at any fixed point tends toward zero;
and it remains zero if the formal limit $\width\rightarrow\infty$
is taken subsequently.
The vortex core is then advected solely by the
Stokes drift $\bubars(0,0)=\bubars=\pmom_1\sliver\bxvec$,
which is now constant across the wavetrain, so that with
$\length\rightarrow\infty$ first in problem~(ii),
(\ref{eq:core-advec-prob-1}) and
(\ref{eq:force-prob-1})
are replaced by

\vspace{-0.35cm}

\begin{equation}
\bubarl(0,0)
~=~
\bubars
~=~
\pmom_1\sliver\bxvec
\label{eq:core-advec-prob-ab}
\end{equation}
and

\vspace{-0.9cm}

\begin{equation}
\bR
~=~
-\Gamma\sliver\pmom_1\sliver\byvec
~.
\label{eq:force-prob-ab}
\end{equation}
Not only the magnitudes but also the signs have changed.
These results hold
for arbitrary $k_0\sliver\radius_0$,
because
in the unrefracted wavetrain
we always have $\bubars=\bpmom=\pmom_1\sliver\bxvec$. \
Notice that (\ref{eq:force-prob-ab}) is equal to the \clvf\
$\bubars\cross\bomegabar=\bubars\cross\bomega_0$
integrated over the vortex core, corresponding to the Iordanskii force
in the quantum fluids literature \citep[e.g.][]{Sonin:1997},
with $\bubars=\bpmom$
corresponding to the phonon current per unit mass.
The relation to the \ab\ effect is discussed in the next two sections.

In any other version of problem~(ii)
there will be two contributions, one from
the \clvf\ and the other from the return part of the Bretherton flow.
In the opposite formal limit, with
the width $\width$ of the wavetrain going to infinity first,
the two contributions cancel.
The return flow is then uniform, and equal and
opposite to the Stokes drift, being diluted only
near the extremities $y\sim\pm\half\width$.
In that formal limit, therefore, the recoil force $\bR$ vanishes.

In all other versions of problem~(ii) there is always
some dilution, making the
return flow weaker than the Stokes drift
and keeping the sign of the recoil opposite to that in problem~(i).
For instance, consider the ``square'' limit
$\length=\width\rightarrow\infty$.
Then
it is readily shown that

\vspace{-0.75cm}

\begin{equation}
\bubarl(0,0)
~=~
\half
\bubars(0,0)
~=~
\half\sliver
\pmom_1\sliver\bxvec
\label{eq:core-advec-prob-square}
\end{equation}
so that in place of (\ref{eq:force-prob-ab}) we have

\vspace{-0.45cm}

\begin{equation}
\bR
~=~
-\half
\Gamma\sliver\pmom_1\sliver\byvec
~.
\label{eq:force-prob-square}
\end{equation}
More generally, the factor $\half$ is replaced by
\,$\{1 -\, 2\sliver\pi^{-1}\Antisliver\lim{\hskip0.1pt}\arctan(\width/\length)\}$.\,

To derive this last result, one can
regard the wide wavetrain as made up of narrow wavetrains each
with $S=\pmom_1\sliver dy$, for constant $\pmom_1$, and then
integrate (\ref{eq:force-prob-1}),
with $Y$ replaced by $y$,
over the whole wavetrain to get
the contribution to $\bR$ from the return flow.
That contribution,
for $\length\rightarrow\infty$ and $\width\rightarrow\infty$
in various ways, is therefore

\vspace{-0.35cm}

\begin{equation}
\frac{\Gamma\pmom_1}{\pi}
\Sliver\lim\int_{-\half\width}^{\half\width}
\frac{\xextent\Sliver dy}{\xextent^2 + y^2}
\Sliver(+\byvec)
\;=\;
\frac{2\Gamma\sliver\pmom_1}{\pi}
\sliver
\lim\sliver\arctan(\width/\length)
\,\Sliver\byvec
~.
\label{eq:arctan-formula}
\end{equation}

In the next
two sections, the foregoing results for problem~(ii) will be cross-checked
against computations of far-field \psm\ fluxes correct to \oaae,
taking account of the refracted wavecrest shapes illustrated in 
figure~\ref{fig:wavecrests-exaggerated}
and their mathematical description (\ref{eq:disloc-wavefield}).
The impulse--\psm\ theorem tells us that the results must agree;
but it is interesting,
nevertheless, not only to carry out the cross-checks but
also to see how the refraction works in more detail,
thereby gaining mechanistic insight, and another view of
the \nil.\footnote{Perhaps
surprisingly, the quantum fluids literature -- going back over the
past fifty years or so -- tends
to ignore many of the points under discussion,
including the \oaa\ mean flow problem, the
distinction between momentum and \psm, and the \nil.
The author has, however, found one big quantum fluids paper \citep{Sonin:1997}
in which the non-interchangeability of limits is
mentioned toward the end of the paper,
almost as an afterthought; see below equation~(83) therein.
Another paper \citep{Wexler:1998} takes a different path but
flags up the dangers of manipulating divergent infinite series.
Some but not all of the \oaa\ effects are discussed in
\citet{Stone:2000b}, while all
are consistently dealt with in
\citet{Guo:2014}, within the Gross--Pitaevskii superfluid model,
but only for problem~(i).
The issues still seem to be surrounded
by controversy, perhaps involving unconscious assumptions
\citep[e.g.][]{McIntyre:2017}
about, for instance, the distinction between particles and quasiparticles.
}

\vspace{-0.31cm}

\section{Wave refraction in problem~(ii): the far field outside the wake}
\label{sec:refrac-outside-wake}

The impulse-\psm\ theorem implies that $\bR$ can be computed as
$-\antisliver\oint \bfpmom\Sliver\cdott\bnvec\Sliver ds$,
correct to \oaae,
for a steady wave~field whose sources and sinks lie
outside the contour of integration.
The unit normal $\bnvec$ is directed outward,
and $\bfpmom$ is the pseudomomentum flux tensor appearing in
(\ref{eq:psm-law}).
We take advantage of the simplicity of the refracted far-field solution
(\ref{eq:disloc-wavefield}), and its compatibility with ray theory and
the group-velocity property (\ref{eq:pmom-flux-gp-vel}),
by letting the contour expand appropriately
as $\length\rightarrow\infty$ and $\width\rightarrow\infty$.
It is convenient to take the contour to be a rectangle with dimensions 
$\length$ by $\width$,
where $\length$ is now to be read as the length of the wavetrain excluding
its source and sink regions,
as they recede to infinity.
Then, correct to \oaae,

\vspace{-0.50cm}

\begin{equation}
\bR
\,=\,
-\lim\antisliver
\oint \bfpmom\Sliver\cdott\bnvec
\Sliver ds
\;=\,\lim
\left(
  \int_{-\tinyhalf\width}^{\tinyhalf\width} \antisliver
  \bfpmom\Sliver\cdott\bxvec
  \Sliver dy\sliver
\bigg|_{x=-\tinyhalf\length}
-
  \int_{-\tinyhalf\width}^{\tinyhalf\width} \antisliver
  \bfpmom\Sliver\cdott\bxvec
  \Sliver dy\sliver
\bigg|_{x=\tinyhalf\length}
\right)
.
\label{eq:pmom-flux-integrals}
\end{equation}
We have used 
(\ref{eq:disloc-wavefield})
and (\ref{eq:pmom-flux-gp-vel})
to neglect the contributions from the
sides of the rectangle parallel to $\bxvec$,
as follows.
For the transverse, $y$ component,
the only relevant component of
(\ref{eq:pmom-flux-gp-vel})
on the sides parallel to $\bxvec$ is \
$\fpmom_{22} =
\byvec\sliver\cdott\antisliver\bpmom\Sliver\,\bCabs\antisliver\cdott\byvec$. \
In the far field, again thanks to ray theory, we have \
$\bpmom \,=\, \bubars \,=\, {\cal A}\sliver\bk$ \
where $\,{\cal A}\,$ is the wave-action per unit mass and where \
$\bk
\,=\,
\bnabla\Phi
\,=\,
k_0\{
\bxvec
-
\frma\sliver\radius_0\sliver\radius^{-1}\sliver\bthetavec
+ O(\frma^2\sliver\radius_0^2\sliver\radius^{-2})
   \}
$, \
from (\ref{eq:disloc-wavefield}) and (\ref{eq:alpha-super-small}) or
from (\ref{eq:rotated-wavenumber}). \
We also have $\wspeed = \wspeed(\radius)
= \wspeed_0\{1 + O(\frma^2\sliver\radius_0^2\sliver\radius^{-2})\}$,
as noted in \S\ref{sec:intro}. \
Denoting $\bpmom\Sliver\cdott\bxvec$ by $\pmom_1$ as before,
and $\bpmom\Sliver\cdott\byvec$ by $\pmom_2$,
we have
$\pmom_2/\pmom_1
= \bk\sliver\cdott\antisliver\byvec/\bk\sliver\cdott\antisliver\bxvec
= O(\frma\sliver\radius_0\sliver\radius^{-1})$. \
From the formula
(\ref{eq:abs-gp-vel}), again noting the cancellation of
leading-order $y$ components,
we have
$\bCabs\antisliver\cdott\byvec = 
O(\frma^2\wspeed_0\sliver\radius_0^2\sliver\radius^{-2})$.
Hence
$\fpmom_{22} = \pmom_1\sliver\wspeed_0$ times
$O(\frma^3\sliver\radius_0^3\sliver\radius^{-3})$.
For the longitudinal, $x$~component, we have
$\fpmom_{12} = \pmom_1\Sliver\bCabs\antisliver\cdott\byvec$
$= \pmom_1\sliver\wspeed_0$ times
$O(\frma^2\sliver\radius_0^2\sliver\radius^{-2})$.
Both make negligible contributions as the rectangle
expands to infinity.

Now it is clear from \S\ref{sec:breth-recoil-2} that $\bR \:||\: \byvec$,
so that $\bR = \nbR\sliver\byvec$, say,
correct to \oaae.  For the sake of brevity, therefore,
we restrict attention from now on to evaluating $\nbR$
from the $y$ component of
(\ref{eq:pmom-flux-integrals}).  We do this
in two stages, to be described in this and the next section.
The first stage is to compute the contribution $\nbRo$ from
outside the wake.
The \nil\ comes from that contribution.  Outside the wake, we can again use
(\ref{eq:disloc-wavefield}), (\ref{eq:rotated-wavenumber}),
(\ref{eq:abs-gp-vel}) and (\ref{eq:pmom-flux-gp-vel}).
At the second stage, in the next section,
we compute $\nbR=\nbRo+\nbRw$, where $\nbRw$ is
the contribution from within the wake.
It will be found
that $\nbRw$ agrees with
(\ref{eq:force-prob-ab}) and that it is
proportional to the \abpj.
The first contribution $\nbRo$ will be found to agree with
(\ref{eq:arctan-formula}) and to
depend solely on the other relevant refraction effect,
the \oeps\ rotation of wavecrests seen in
figure~\ref{fig:wavecrests-exaggerated} and expressed by the
$\bthetavec$ term in $\bk$. \
From here on we denote $\bfpmom$ evaluated from
(\ref{eq:disloc-wavefield}), (\ref{eq:rotated-wavenumber}),
(\ref{eq:abs-gp-vel}) and (\ref{eq:pmom-flux-gp-vel})
by $\bfpmomo$, so that

\vspace{-0.5cm}

\begin{equation}
\nbRo
\antisliver=\lim\byvec\Sliver\cdott\!
\left(
  \int_{-\tinyhalf\width}^{\tinyhalf\width} \antisliver
  \bfpmomo\cdott\bxvec
  \Sliver dy\sliver
\bigg|_{x=-\tinyhalf\length}
\!\!-\!\!
  \int_{-\tinyhalf\width}^{0-} \antisliver
  \bfpmomo\cdott\bxvec
  \Sliver dy\sliver
\bigg|_{x=\tinyhalf\length}
\!\!-\!\!
  \int_{0+}^{\tinyhalf\width} \antisliver
  \bfpmomo\cdott\bxvec
  \Sliver dy\sliver
\bigg|_{x=\tinyhalf\length}
\right)
\label{eq:pmom-flux-integrals-outside}
\end{equation}
which, in the limit, correctly excludes the wake contribution
because of the relative narrowness
of the wake,
whose width $\wakewidth\ll\width$, as detailed in the next section.

We can evaluate
$\byvec\Sliver\cdott\bfpmomo\cdott\bxvec
= \pmom_2\Sliver\bCabs\cdott\bxvec$
from
$\bCabs\cdott\bxvec = \wspeed_0 \{
1 + O(\frma\sliver\radius_0\sliver\radius^{-1})
\}$
and
$\pmom_2
= \pmom_1\bk\sliver\cdott\antisliver\byvec/\bk\sliver\cdott\antisliver\bxvec
=-\pmom_1\{
\frma\sliver\radius_0\sliver\radius^{-1}\sliver\bthetavec\sliver\cdott\antisliver\byvec
+ O(\frma^2\sliver\radius_0^2\sliver\radius^{-2})
\}
= \pmom_1 k_0^{-1}\{
\partial\sliver\Phi/\partial y + O(\frma^2\sliver\radius_0^2\sliver\radius^{-2})
\}$
with $\Phi$ as in (\ref{eq:disloc-wavefield}), and
where $\pmom_1$ can now be read as
the incident \psm, neglecting refraction, that is,
$\pmom_1=$ const.\ as in \S\S\ref{sec:breth-recoil-1}--\ref{sec:breth-recoil-2}.
Correct to \oaae, therefore, the first integral on the right of
(\ref{eq:pmom-flux-integrals-outside})
reduces to
\,$
\wspeed_0\sliver\pmom_1 k_0^{-1}
\lim\int_{-\alpha\thetatilde}^{+\alpha\thetatilde}
\Sliver d\Phi\sliver
=
\wspeed_0\sliver\pmom_1 k_0^{-1}\lim\sliver(2\alpha\sliver\thetatilde)
=
(\Gamma\sliver\pmom_1/\pi)
\sliver\lim\sliver\thetatilde$,
where $\thetatilde=\arctan(\width/\length) > 0$. \
The noninterchangeability of the limits
$\length\rightarrow\infty$ and $\width\rightarrow\infty$
is now evident.

At each fixed $y$, and correct to \oaae,
$\pmom_2$ is an odd function of $x$, and
$\byvec\Sliver\cdott\bfpmomo\cdott\bxvec$ also.
Therefore the second and third integrals in
(\ref{eq:pmom-flux-integrals-outside})
add up to a contribution equal to that from the first integral,
so that altogether

\vspace{-0.35cm}

\begin{equation}
\nbRo
\;=\;
\frac{2\Gamma\sliver\pmom_1}{\pi}
\sliver\lim\sliver\arctan(\width/\length)
\label{eq:R-outside}
\end{equation}
correct to \oaae,
in agreement with (\ref{eq:arctan-formula}).
Notice incidentally that
problem~(i) now appears as a trivial
variant of the above, obtained by selecting appropriate subsets of rays
in cases where ray theory is valid all along the wavetrain.

\vspace{-0.3cm}

\section{Wave refraction in problem~(ii): the far field within the wake}
\label{sec:refrac-within-wake}

To complete the work on problem~(ii) we need to
evaluate the remaining contribution to the $y$ component of
(\ref{eq:pmom-flux-integrals}),

\vspace{-0.45cm}

\begin{equation}
\nbRw
\;=\;
-\lim
\int_{\rm wake} \antisliver
  \byvec\Sliver\cdott\bfpmomw\cdott\bxvec
\Sliver dy\sliver
\bigg|_{x=\tinyhalf\length}
,
\label{eq:pmom-flux-wake}
\end{equation}
and to verify that it agrees with (\ref{eq:force-prob-ab}).
Here $\bfpmomw$ stands for $\bfpmom$ within the wake.

We evaluate (\ref{eq:pmom-flux-wake}) in two cases that are
analytically tractable,
$k_0\sliver\radius_0 \ll 1$ and
$k_0\sliver\radius_0 \gg 1$.  In the second case we
use ray tracing across the vortex core, and in the first we draw on
the work of
\citet[][hereafter FLS]{Ford:1999}, who carried out
a careful and thorough asymptotic analysis
of weak refraction and scattering
in that case, building on earlier contributions including that of
\citet{Sakov:1993}; see also \citet{Belyaev:2008}.

\vspace{-0.2cm}

\subsection{The long-wave case $k_0\sliver\radius_0 \ll 1$}
\label{subsec:longwave}

FLS's results are
subject to
a severe restriction on the range of $\alpha$ values
for which they are valid.
From (\ref{eq:alpha-super-small}) we see that
$\alpha$ is now the product of two small quantities
$k_0\sliver\radius_0$ and $\frma$.
The results are nevertheless attractive for our purposes because, when valid,
they show that the wake has
a simple Fresnel-diffractive structure with width-scale
$\wakewidth \sim k_0^{-1/2}\length^{1/2} \gg k_0^{-1}$,
and angular scale asymptotically zero, as $\length\rightarrow\infty$.
Across the wake there is a smooth phase transition
such that the ray-theoretic formulae used in \S\ref{sec:refrac-outside-wake}
still hold, including $\bpmom \propto \bk$ and
$\pmom_2 = \pmom_1 k_0^{-1} \partial\Phi/\partial y$,
where $\Phi$ now denotes the phase within the wake, as distinct from the
phase given by (\ref{eq:disloc-wavefield}).
The wave~field within the wake still has the form
$\envel\exp(\imag\sliver\Phi)$, where the amplitude
$\envel$ is still real and constant to sufficient accuracy
across the wake, with relative error \oeps,
and where the phase $\Phi$ increases by $2\pi\alpha$
going anticlockwise across the wake.
Therefore we can evaluate
$\int\byvec\sliver\cdott\sliver\bfpmomw\sliver\cdott\sliver\bxvec\Sliver dy$\,
across the wake as 
\,$\int \wspeed_0\sliver\pmom_2\Sliver dy =
\wspeed_0\sliver\pmom_1 k_0^{-1}\int\!d\Phi =
2\pi\alpha \wspeed_0\sliver\pmom_1 k_0^{-1} = \Gamma\pmom_1$,\,
verifying that (\ref{eq:pmom-flux-wake}) with its minus sign
does agree with (\ref{eq:force-prob-ab}).

FLS's solution also contains a Born-scattering term with amplitude
$O(\radius^{-1/2})$, which however contributes nothing.
The magnitude $O(\radius^{-1})$ of its \psm\ flux
makes it potentially able to contribute to
$-\antisliver\oint \bfpmom\Sliver\cdott\bnvec\Sliver ds$.
But for our rectangular integration contour the
contribution to $\byvec\sliver\cdott\bfpmom\sliver\cdott\bnvec$
is an odd function of $y$, which integrates to zero.

The reader who wishes to check the foregoing against FLS in more detail
may find the following notes useful.
Outside the wake, the leading far-field term
in FLS's solution agrees with the foregoing for any
$\alpha\ll 1$ because we then have, from
(\ref{eq:disloc-wavefield}),

\vspace{-0.5cm}

\begin{eqnarray}
\exp(\imag\sliver\Phi)
~=~& \hspace{-1.3cm}
\exp(-\imag \alpha\theta)
\exp\{\imag ( k_0(x-\wspeed_0 t) + \mbox{\rm const.} )\}\nonumber\\
~=~&
(1 - \imag \alpha\theta)
\exp\{\imag ( k_0(x-\wspeed_0 t) + \mbox{\rm const.} )\}
+ O(\alpha^2)
~.
\label{eq:disloc-wavefield-small-alpha}
\end{eqnarray}
This agrees with FLS's (2.14), (2.20) and the leading term in their (5.7),
after allowing for their different definition of $\theta$
and remembering that our $\theta$ jumps from $+\pi$ to $-\pi$,
going anticlockwise across the positive $x$~axis.
In their dimensionless notation, our $\alpha$
is written as $M^2\Gamma\omega/2\pi$,
where
their $M$ is our $\frma$
and their $\Gamma\omega/2\pi$ is our $k_0\radius_0/\frma$,
all taken positive.  The
Born scattering term is
the next, $O(\radius^{-1/2})$ term in their (5.7),
with outgoing waves
$\propto\radius^{-1/2}\exp\{\imag k_0(\radius-\wspeed_0 t)\}$.
The Fresnel wake
is described by their (5.12).
The phase transition at fixed $x$
is given by the sum $C+S$ of two
real-valued Fresnel integrals, suitably scaled,
and is therefore an odd function of
$y/\wakewidth$,
with $\Phi$ asymptoting toward $\pm\pi\alpha+\mbox{const.}$
with gentle oscillations on the scale $\wakewidth$.
The factor $\exp(-\imag\eta^2)$ in FSL's (5.11),
where $\eta^2 = \half (y/\wakewidth)^2$,
converts a Born-like factor
$\exp\{\imag k_0(\radius-\wspeed_0 t)\}$ into a plane-wave factor 
$\exp\{\imag k_0(x      -\wspeed_0 t)\}$,
to sufficient accuracy within the wake,
matching up with our (\ref{eq:disloc-wavefield-small-alpha}).

\citet{Belyaev:2008} reconsider FLS using a different solution technique,
that of \citet{Aharonov:1959} and \citet{Berry:1980}.
They also
discuss the conceptual issue of whether
(\ref{eq:disloc-wavefield}) above can usefully be regarded as a
plane wave outside the Fresnel wake, in the limit 
$\radius\rightarrow\infty$.
However, the wave~field properties required by the foregoing analysis of
the \psm\ budget are unaffected.  Those properties are, most crucially,
the validity of (\ref{eq:disloc-wavefield}) outside the wake,
the phase continuity across the wake, and
the $y$-antisymmetry of the Born contribution
to $\byvec\sliver\cdott\bfpmom\sliver\cdott\bnvec$.
And we note again that the analysis is independently confirmed by
the end-to-end cross-check from \S\S\ref{sec:imp-psm}--\ref{sec:breth-recoil-2}.

\vspace{-0.2cm}

\subsection{The short-wave case $k_0\sliver\radius_0 \gg 1$}
\label{subsec:shortwave}

We now evaluate (\ref{eq:pmom-flux-wake}) in
the case $k\sliver\radius_0 \gg 1$, using ray theory.
Attention is restricted to the
simplest case, the Rankine vortex model in which
the core is in solid rotation, with constant angular velocity
$\Omega = \half|\bomega_0| = |\bu_0(r)|/r = \Gamma/2\pi\radius_0^2$,
taken positive, i.e.\ anticlockwise, for definiteness,
as in the figures.

Correct to \oeps\
the rays outside the core and its lee are straight,
as already remarked, with absolute group velocities $\bCabs$
parallel to $\bxvec$,
 but slightly rotated wavecrests, as seen in
figure~\ref{fig:wavecrests-exaggerated}.
Inside the core, the ray-tracing
equations, e.g.\ (2.14) of BM03, verify what is obvious
from rotating the reference frame while keeping the same
intrinsic phase and group velocity
$\wspeed = \wspeed_0\{1 + O(\frma^2)\}$,
namely that the wavenumber vector $\bk$ rotates with angular velocity
$\Omega$
as a ray point crosses the core at
velocity $\wspeed_0\sliver\bxvec + O(\frma)$.
Thus the rays bend slightly to the left as they cross the core into its lee,
where they are straight again but no longer quite
parallel to the $x$~axis.  In fact
the group velocity $\bCabs$ rotates twice as fast as $\bk$,
with angular velocity $2\Omega$,
because of the changing \oeps\ contribution from the $y$ component of
$\bu_0 = \Omega(-y,x)$ in (\ref{eq:abs-gp-vel})
as the ray point crosses the core.
This and the straightness of the rays outside the core
are special cases of
the curl-curvature formula mentioned below
(\ref{eq:abs-gp-vel}).
The formula implies generally that, correct to \oeps,
the group velocity vector
rotates with angular velocity $\bomega_0=\bnabla\cross\bu_0$.

\begin{figure}
\centering
\vspace{-0.2cm}
\includegraphics[width=11 cm, viewport=0 0 800 195,
clip]{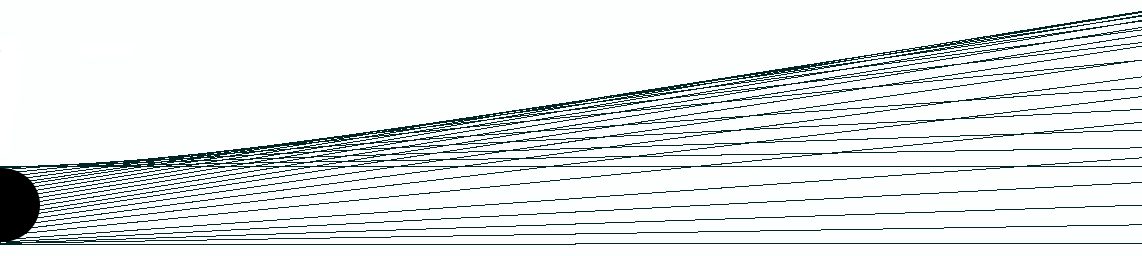}
\vspace{0.1cm}
\caption{Rays emerging from the vortex core and forming a wake with a caustic,
computed correct to \oeps\ from the ray-tracing equations.
The \oeps\ deflections are exaggerated in the plot.
The rays entering the core, not shown, are initially parallel to the
$x$~axis at $y/\radius_0 = 0$, $\pm0.1$, $\pm0.2$, $\pm0.3$, $\pm0.4$,
$\pm0.5$, $\pm0.6$, $\pm0.7$, $\pm0.8$, $\pm0.9$,
$\pm0.95$, $\pm0.98$, $\pm0.995$, $\pm1$.
}
\vspace{-0.35cm}
\label{fig:caustic}
\end{figure}

For weak refraction the ray undergoing the greatest deflection
is that crossing the widest part of the core,
at $y=0$.  The rays from $-\radius_0 < y < 0$ therefore splay out
slightly, while those from
$0< y < \radius_0$ cross one another and form a caustic,
extending slightly outside the tangent line $y=+\radius_0$,
as shown in figure~\ref{fig:caustic} with the deflections exaggerated.
A full analysis is beyond our scope here;
to evaluate (\ref{eq:pmom-flux-wake}) we will
simply add up the leading-order \psm\ fluxes
$\byvec\sliver\cdott\sliver\bfpmomw\sliver\cdott\sliver\bxvec =
\wspeed_0\sliver\pmom_2$ as if
carried by each ray independently.  This assumes that the refraction term
in the \oaae\ \psm\ law (\ref{eq:psm-law}) works in the same way,
to leading order at least, whether or not the rays go through a caustic.

The treatment of the ray deflections as small of order $\frma$,
and the neglect of diffractive effects,
becomes delicate and very restrictive when
combined with the formal limit \mbox{$\length\rightarrow\infty$}.
It will nevertheless yield the correct result,
in agreement with (\ref{eq:force-prob-ab}), as will now be shown.
The agreement will also lend support to our assumption about
(\ref{eq:psm-law}) and caustics.

The following is a shortcut to the results from the ray-tracing equations
shown in figure~\ref{fig:caustic}, using the rotation-rate of $\bk$
already mentioned.
A ray point entering the core at $y=-\radius_0\sin\theta$,
say, for some fixed $\theta$ with $|\theta| < \pi/2$,
already has a wavenumber with nonvanishing $y$ component
$\bk\Sliver\cdott\byvec =
-\alpha\radius_0^{-1}\bthetavec\Sliver\cdott\byvec =
+\alpha\radius_0^{-1}\cos\theta$,
from (\ref{eq:disloc-wavefield}) or (\ref{eq:rotated-wavenumber}) giving,
at that location,
$\byvec\sliver\cdott\sliver\bfpmomw\sliver\cdott\sliver\bxvec =
\wspeed_0\sliver\pmom_2 =
\wspeed_0\sliver\pmom_1 k_0^{-1}\alpha\radius_0^{-1}\cos\theta =
\pmom_1\Gamma(2\pi\radius_0)^{-1}\!\cos\theta =
\pmom_1\Omega\sliver\radius_0\cos\theta
$. \
As the ray point crosses the core,
over a distance
$2\radius_0\cos\theta$
and taking a time
$2\sliver\wspeed_0^{-1}\radius_0\cos\theta$, the vectors
$\bk$ and $\bpmom$ rotate with angular velocity $\Omega$,
so that $\wspeed_0\sliver\pmom_2$ increases by a further small amount
$2\pmom_1\Omega\sliver\radius_0\cos\theta$ (equal to
$\wspeed_0\sliver\pmom_1$ times the net rotation angle
$2\sliver\wspeed_0^{-1}\Omega\sliver\radius_0\cos\theta$),
and again by a further small amount
$\pmom_1\Omega\sliver\radius_0\cos\theta$
after exiting the core and reaching sufficiently large $x>0$.
This last increment is the same increment as in
(\ref{eq:disloc-wavefield}) between the
far edge of the core and $x\rightarrow\infty$.
However, it is to be added to
the new far-core-edge value
$3\pmom_1\Omega\sliver\radius_0\cos\theta$ rather than to the
original value
\,$-\pmom_1\Omega\sliver\radius_0\cos\theta$
implied by (\ref{eq:disloc-wavefield}) and (\ref{eq:rotated-wavenumber}).
With our ray still nearly parallel to the $x$~axis
after exiting the core, we are using the fact that
the \oeps\ \emph{rates of change}
of $\bk\sliver\cdott\byvec$ outside the core are the same as those implied
by (\ref{eq:disloc-wavefield}), (\ref{eq:rotated-wavenumber})
and (\ref{eq:abs-gp-vel}).

So, adding all the contributions just noted, we have that
the ray has a total end-to-end change 
$
 \pmom_1\Omega\sliver\radius_0\cos\theta+
2\pmom_1\Omega\sliver\radius_0\cos\theta+
 \pmom_1\Omega\sliver\radius_0\cos\theta=
4\pmom_1\Omega\sliver\radius_0\cos\theta
$,
in $\wspeed_0\sliver\pmom_2$,
corresponding to an end-to-end deflection angle
$\beta$, say, $=\pmom_2/\pmom_1|_{x\rightarrow\infty} =
4\wspeed_0^{-1}\Omega\sliver\radius_0\cos\theta=4\frma\cos\theta$, with
maximum value $\sliver4\frma\,$, in an anticlockwise sense.
Integrating the change in $\wspeed_0\sliver\pmom_2$
over all the rays that cross the core, from
$y=-\radius_0$ to
$y= \radius_0$,
noting that $dy=-\radius_0\cos\theta\Sliver d\theta$ and that
$\int_{-\pi/2}^{\pi/2}\antisliver\cos^2\Antisliver\theta\Sliver d\theta
= \pi/2$,
we find from (\ref{eq:pmom-flux-wake}) with due care over signs that
$\nbRw =
-2\pi\pmom_1\Omega\sliver\radius_0^2 = -\Gamma\pmom_1$,
agreeing with (\ref{eq:force-prob-ab}) as expected.

To summarize, (\ref{eq:pmom-flux-integrals}) gives us, both for
$k_0\sliver\radius_0 \ll 1$ and for
$k_0\sliver\radius_0 \gg 1$,

\vspace{-0.3cm}

\begin{equation}
\nbR\;=\;\nbRw+\,\nbRo
\;=\;
-\left\{
1 -
\frac{2}{\pi}\sliver\lim\sliver
\arctan\left(\frac{\width}{\length}\right)
\right\}
\Gamma\sliver\pmom_1
\label{eq:Rw-plus-Ro}
\end{equation}
for arbitrary limiting values of $\width/\length$,
agreeing with the independent derivations in
\S\S\ref{sec:imp-psm}--\ref{sec:breth-recoil-2}.
Furthermore, recalling that those independent derivations
are valid for arbitrary $k_0\sliver\radius_0$,
we see also that
(\ref{eq:Rw-plus-Ro}) must be a result far more robust than
is suggested by the delicacy of the flux computations for
$k_0\sliver\radius_0 \ll 1$ and
$k_0\sliver\radius_0 \gg 1$.
On the other hand,
everything still
depends on the smallness of $\frma$.
Numerical solutions for
cases of stronger refraction
show wakes oriented at substantial angles away from the $x$ axis;
see for instance figures~2--3 of \citet{Coste:1999}.
And it is still an open question as to what might or might not
replace the impulse--\psm\ theorem for arbitrary $\frma$.

\vspace{-0.45cm}

\section{Problem~(iii)}
\label{sec:prob3}

The main reason for being interested in
this rapidly-rotating version of problem~(i),
in which the waves are deep-water gravity waves, is
the existence of the Ursell anti-Stokes flow.
This is
 an Eulerian-mean flow $\bubar$ that largely cancels the
strongly $z$-dependent Stokes drift $\bubars$ of
the waves.  Indeed the cancellation is exact,
for finite-amplitude waves,
when the wave~field is exactly steady and exactly homogeneous
across an infinite $xy$ domain \citep{Ursell:1950,Pollard:1970}.
\ (In a \te\ starting with irrotational waves,
in such a domain,
the mean flow undergoes a free inertial oscillation
about the anti-Stokes state.  It is sometimes forgotten that
this \te\ was clearly analysed and understood in Ursell's
pioneering work.) \
In problem~(iii), however,
contrary to what might at first be thought, the anti-Stokes flow
fails to suppress remote recoil.

We consider an unstratified rapidly-rotating system
of finite depth $H$ under gravity $-g\bzvec$, so that the $z$ direction is
vertically upward.
The vector Coriolis parameter $\bfcoriolis$ is
parallel to $\bzvec$.
The waves
have a wavenumber $k=|\bk|$ that is large enough to
make $\exp(-kH)$ negligible.
The intrinsic wave frequency
$k\wspeed = (gk)^{1/2} \gg \fcoriolis = |\bfcoriolis|$,
so that rotation affects the wave dynamics only weakly.

In addition to $\amp$ and $\frma$ we now have another small parameter,
the mean-flow Rossby number

\vspace{-0.7cm}

\begin{equation}
\Ro
~=~
U/\antisliver\fcoriolis\radius_0
~\ll~
1
~,
\label{eq:rossby-number}
\end{equation}

\smallskip
\noindent
whose smallness will bring in
the Taylor--Proudman effect
and give us \qg\ mean-flow dynamics.
As before, the velocity scale $U$ will be
taken as the velocity of the vortex flow at the edge of the core,
$\radius = \radius_0$.
The core will be defined by nonvanishing
\qg\ potential vorticity,
$(\nablahsq - \LD^{-2})\psitilde_0$\, in the notation of \S\ref{sec:eqns}.

The anti-Stokes flow can be regarded as a consequence of
the Taylor--Proudman effect
together with
the exact advection property
expressed by the mean vorticity equations
(\ref{eq:3d-exact-mean-vort}) and (\ref{eq:3d-exact-mean-vort-cdot-form}).
This is no more than a rephrasing of Ursell's original argument,
putting it within the GLM framework.  Focusing on
the present case $\Ro \ll 1$, we can regard
inertia waves as fast waves with a strong Coriolis restoring force.
The Taylor--Proudman effect arises from the corresponding stiffness
of the lines of absolute vorticity
$\bfcoriolis + \bomegatilde$, which
must tend to stay vertical, on average at least.  In particular, they cannot
be continually sheared over by the mean flow;
and for this purpose the mean flow is
the Lagrangian-mean flow $\bubarl$, as equations
(\ref{eq:3d-exact-mean-vort}) and (\ref{eq:3d-exact-mean-vort-cdot-form})
make clear.
A Lagrangian-mean flow without vertical shear is a Stokes drift plus
an Eulerian-mean anti-Stokes flow, plus an additional contribution
that is independent of $z$ -- in this case the vortex flow plus the
Bretherton flow
that mediates remote recoil.

To tackle problem~(iii) we must first derive (\ref{eq:for-prob-3}).
The starting point is
the vertical component of~(\ref{eq:3d-exact-mean-vort-cdot-form}).
Writing $\fcoriolis + \sliver\omegatilde$
for the vertical component of
$\bfcoriolis + \sliver\bomegatilde$, and $\wbarl$
for the vertical component of $\bubarl$, we have

\vspace{-0.35cm}

\begin{equation}
\frac{\Dbarl\omegatilde}{Dt}
~+~
(\fcoriolis + \sliver\omegatilde)\bnabla\cdott\bubarl
~=~
(\bfcoriolis + \sliver\bomegatilde)\sliver\cdott\antisliver\bnabla\wbarl
\label{eq:3d-exact-mean-vort-cdot-form-vert-cpt}
\end{equation}
exactly. \ Upon cancelling a pair of terms in
$\partial\wbarl\antisliver/\partial z$, this reduces to

\vspace{-0.25cm}

\begin{equation}
\frac{\Dbarl\omegatilde}{Dt}
~+~
(\fcoriolis + \sliver\omegatilde)\bnablah\cdott\bubarlh
~=~
\bomegatildeh\sliver\cdott\antisliver\bnablah\wbarl
.
\label{eq:3d-exact-mean-vort-cdot-form-vert-cpt-divh}
\end{equation}
As before, suffix {\small H} denotes horizontal projection.
Writing $\omegatilde=\omega_0+\omegatildeb$ and $\bubarl=\bu_0+\bubarlb$
where the vortex-only contributions $\omega_0$ and $\bu_0$
are $z$-independent, with $\bu_0$ horizontal,
and the wave-induced contributions $\omegatildeb$ and $\bubarlb$ are \oaa,
we note that in the first term on the left the contribution
$\bubarlb\sliver\cdott\Antisliver\bnabla\Sliver\omegatildeb \,=\,
\bubarlb\sliver\cdott\Antisliver\bnablah\Sliver\omegatildeb \,+\,
\wbarl\sliver\partial\Sliver\omegatildeb\antisliver/\partial z \,=\,$
\oaaaa\ and is therefore negligible.
(We need not restrict $\frma$ at this stage.)
There are two further such \oaaaa\ contributions, namely
the \rhs, and on the left
$\omegatildeb\sliver\bnablah\cdott\bubarlh =
 \omegatildeb\sliver\bnablah\cdott\bubarlb$
since $\bnablah\cdott\bu_0=0$. \
The \oaa\ contribution
$\omega_0\Sliver\bnablah\cdott\bubarlh=
 \omega_0\Sliver\bnablah\cdott\bubarlb$
is also negligible, against
$\fcoriolis\Sliver\bnablah\cdott\bubarlh$, because of the
smallness of $\Ro$. \
After neglecting
all these contributions we can take
the vertical
average of (\ref{eq:3d-exact-mean-vort-cdot-form-vert-cpt-divh}),
using the $z$-independence of $\bu_0$ and $\bomega_0$.
Denoting vertical averages by angle brackets as before and noting that
$\bu_0\sliver\cdott\Antisliver\bnablah\Sliver\omega_0 = 0$ we get

\vspace{-0.3cm}

\begin{equation}
\left(
\frac{\partial}{\partial t} + \bu_0\sliver\cdott\Antisliver\bnablah
\right)
\langle\omegatildeb\rangle
\;+\;
\langle\bubarlb\rangle\sliver\cdott\Antisliver\bnablah\Sliver\omega_0
\;+\;
\fcoriolis\Sliver\bnablah\cdott\langle\bubarlh\rangle
~=~
0
\label{eq:3d-mean-vort-cdot-form-vert-cpt-divh-integd-expanded}
\end{equation}
or, written more compactly, again with negligible error \oaaaa,

\vspace{-0.35cm}

\begin{equation}
\frac{\Dbarlh\langle\omegatilde\rangle}{Dt}
~+~
\fcoriolis\Sliver\bnablah\cdott\langle\bubarlh\rangle
~=~
0~,
\label{eq:3d-mean-vort-cdot-form-vert-cpt-divh-integd}
\end{equation}
where we have defined
$\Dbarlh/Dt = \partial/\partial t +
\langle\bubarlh\rangle\sliver\cdott\Antisliver\bnablah$.
\ In a closely similar way, the vertical average of the
three-dimensional mass-conservation equation,
B14 equation~(10.47),
simplifies to a vertically-averaged version of
(\ref{eq:mean-mass-conservation}),

\vspace{-0.35cm}

\begin{equation}
\frac{\Dbarlh\htilde}{Dt}
~+~
\htilde\Sliver\bnablah\cdott\langle\bubarlh\rangle
~=~
0
~,
\label{eq:mean-mass-conservation-depth-avd}
\end{equation}
again with negligible error \oaaaa.
As before, the mean layer depth $\htilde=\htilde(x,y,t)$ is defined
such that $\rho\sliver\htilde\Sliver dxdy$ is the areal mass element,
where $\rho$ is the constant mass density.
\ Elimination of $\bnablah\cdott\langle\bubarlh\rangle$ between
(\ref{eq:3d-mean-vort-cdot-form-vert-cpt-divh-integd}) and
(\ref{eq:mean-mass-conservation-depth-avd})
gives us that
$\langle\omegatilde\rangle - \fcoriolis\ln\htilde$, plus an
arbitrary additive constant, is a material invariant
under advection by $\langle\bubarlh\rangle$\antisliver.\,
Fractional changes in $\htilde$ are small,
$(\htilde - H)/H=O(\Ro)$,
and so taking the additive constant to be $\fcoriolis\ln H$
and using \ $\ln\htilde - \ln H = \ln(\htilde/H) = 
\ln\{(\htilde - H + H)/H\} = (\htilde - H)/H + O(\Ro^2)$, we get

\vspace{-0.25cm}

\begin{equation}
\frac{\Dbarlh\qtilde}{Dt}
~=~
0
\label{eq:pv-cons}
\end{equation}
where

\vspace{-0.65cm}

\begin{equation}
\qtilde
~=~
\qtilde(x,y,t)
~=~
\langle\omegatilde\rangle
-
\frac{\fcoriolis}{H}
(\htilde - H)
~,
\label{eq:qgpv-def}
\end{equation}
which is the appropriate form of the \qg\ potential vorticity of the mean flow.
In any \te\ in which the waves are switched on after the vortex is established,
(\ref{eq:pv-cons}) implies that
the $\qtilde$ field is unchanged by the presence of the waves,
apart from the advection of the vortex core by the Bretherton flow.
See also Appendix~\ref{sec:appb}.
So in problem~(iii) we have $\qtilde=q_0$ where
$q_0$ is the potential vorticity of the vortex alone.

The final step in deriving (\ref{eq:for-prob-3}) is
to make explicit use of hydrostatic and geostrophic balance.
Some delicate
scale analysis is involved at this stage.
The full details are given in Appendix~\ref{sec:appc},
in which the key points are as follows.  Hydrostatic balance,
meaning the overall balance for a complete fluid column, implies that
horizontal pressure gradients on the bottom,
underneath the wavetrain, are given by
$\rho\sliver g\bnablah\htilde$, again because
$\rho\sliver\htilde\Sliver dxdy$ is the areal mass element.
Geostrophic balance then gives (\ref{eq:streamfunctiondef}) with
$\psitilde = g(\htilde-H)/\fcoriolis$.
Then $\qtilde = q_0$ together with
(\ref{eq:exact-mean-vorticity}) and (\ref{eq:bstreamfunctiondef})
gives (\ref{eq:for-prob-3}).\;
The Taylor--Proudman effect extends the geostrophic relation upward
into the wavetrain; $\langle\bubarlh\rangle = \bubarlh$.\;
Radiation stresses within the wavetrain
cannot break the overall hydrostatic balance because
such stresses have no foothold on the bottom boundary,
in virtue of our assumption that $\exp(-kH)$ is negligibly small.
That allows us to neglect the net vertical, radiation-stress-induced
external force on the fluid column --
in contrast, it should be noted, with the situation of
figure~\ref{fig:bretherton}.
For further comments see Appendix~\ref{sec:appc}. \
In Appendix~\ref{sec:appc}
we also note that the exact wave solution of \citet{Pollard:1970}
provides some useful cross-checks.

With (\ref{eq:for-prob-3}) in place,
we can now invoke the impulse-\psm\ theorem
to assert that recoil forces can be computed either from
Bretherton flows correct to \oaaeo\ or from net \psm\ fluxes correct to \oaae.
In the remainder of this section we carry out both computations,
in the case of a small vortex core with $\radius_0\ll\LD$,
providing mechanistic insight as
well as an end-to-end cross-check on
our derivation of (\ref{eq:for-prob-3}).

First consider
the Bretherton flow.  Because it satisfies (\ref{eq:for-prob-3}),
it decays sideways like $\exp(-|y|/\LD)$, on the fixed length-scale $\LD$.
Therefore there is no dilution effect like that in problem~(i).
With $\length\rightarrow\infty$,
and with a narrow wavetrain for which $\width\ll \LD$ and $\width\ll Y$,
in the notation of \S\ref{sec:breth-recoil-1},
we have, for $|y-Y| > \width$, outside the unrefracted wavetrain,
with $Y$ the distance to the vortex core,

\vspace{-0.35cm}

\begin{equation}
\bubarlb(x,\,y)
~=~
(S/2\LD) \exp(-|y-Y|/\LD)\sliver(-\bxvec)
\label{eq:core-advec-prob-3}
\end{equation}
where $S$ is still defined by (\ref{eq:mass-flow-prob-1})
but with vertical averaging understood.\; So,
for our small vortex core with $\radius_0\ll\LD$,
carried bodily by the $z$-independent Bretherton flow,
we take $y = 0$ in (\ref{eq:core-advec-prob-3}) to get

\vspace{-0.50cm}

\begin{equation}
\bR
~=~
(\Gamma S/2\LD)
\exp(-|Y|/\LD)\sliver(+\byvec)
~,
\label{eq:force-prob-3}
\end{equation}
with $\Gamma$ evaluated
at the edge of the core.  The signs
are the same as those in problem~(i).

Second, we compute $\bR$ from the \oaae\ \psm\ flux
$\fpmom_{21} = \byvec\sliver\cdott\sliver\bfpmom\sliver\cdott\sliver\bxvec$,
using ray theory.
As in \S\ref{subsec:shortwave},
the rays start exactly parallel to the $x$~axis, with
$\pmom_2\rightarrow0$ as $x\rightarrow-\infty$,
and finish after bending slightly, through
an \oeps\ end-to-end deflection angle
$\beta=\pmom_2/\pmom_1|_{x\rightarrow\infty}$. \
The vortex flow has velocity
$\bu_0(\radius) = \bthetavec\Sliver\partial\psi_0/\partial r$, say, where
the \qg\ streamfunction $\psi_0$ satisfies
$(\nablasq - \LD^{-2})\sliver\psi_0=0$ outside the core.
Defining
$\Radius=\radius/\LD$,
we have

\vspace{-0.7cm}

\begin{equation}
\psi_0
~=\;
-
\frac{\Gamma}{2\pi}
\Sliver
K_0(\Radius)
\qquad \mbox{outside the core}
,
\label{eq:sw-vortex}
\end{equation}
where $K_0(\Radius)$ is the
modified Bessel function
asymptoting to $(\pi/2\Radius)^{1/2}\exp(-\Radius)$
for \mbox{$\Radius\gg1$}
and to $-\ln(\Radius)$
for $\Radius\ll1$, near the core.
The Kelvin circulation $\Gamma$ is again
defined to be the circulation at the core edge $\radius=\radius_0$,
namely $\pm2\pi\radius_0\sliver|\bu_0(\radius_0)|=\pm2\pi\radius_0\sliver U$,
with positive sign
when the vortex is cyclonic as
in the figures.
For $\radius>\radius_0$ the circulation
is not constant, but
decays exponentially like $\Radius^{1/2}\exp(-\Radius)$.

To verify agreement with 
(\ref{eq:force-prob-3})
we need only calculate $\beta$.
The curl-curvature formula
tells us that
$\beta$ is nonzero at \oeps,
because the relative vorticity $\nablasq\psi_0=\LD^{-2}\psi_0$
is nonzero outside the core.
A cyclonic vortex core is surrounded by anticyclonic vorticity
and the rays therefore bend to the right, rather than to the left as
in \S\ref{sec:refrac-within-wake}, so that
$\sgn\sliver\beta=-\sgn\Sliver\Gamma$.
Notice incidentally that there will no longer be any far-field
subtleties, or issues with noninterchangeable limits,
thanks to the exponential decay of $\psi_0$.
Another effect of that decay is that the right-bending rays
must splay out slightly when they pass to the left of the vortex,
but cross one another and form a caustic when to the right.
The presence or absence of a caustic makes no difference to the results.

The deep water waves in problem~(iii)
have intrinsic frequency $k\wspeed = (gk)^{1/2}$ and
intrinsic group velocity $\bC = C\bk/k$ where
$C = \half \wspeed = \half(g/k)^{1/2}$. \
The absolute group velocity $\bCabs = \half\wspeed_0\sliver\bxvec +$ \oeps,
with \oeps\ contributions coming both from $\bu_0$ and from
refractive changes in wavenumber $\bk$.
Following a ray point moving
at speed $\half \wspeed_0 +$ \oeps,
the curl-curvature formula says
that the direction of $\bCabs$ rotates clockwise away from the $x$ direction
at an angular velocity equal to the (negative)
relative vorticity $\nablasq\psi_0=\LD^{-2}\psi_0$.
So for weak refraction
we have

\vspace{-0.35cm}

\begin{equation}
\beta
~=~
(\half \wspeed_0)^{-1}
\int_{-\infty}^\infty
\LD^{-2}\sliver\psi_0(x,Y)
\Sliver
dx
~=\;
-\Sliver
\frac{\Gamma\;}{\pi \wspeed_0\LD^2}
\int_{-\infty}^\infty
K_0\{(x^2+Y^2)^{1/2}/\LD\}
\sliver
dx
~.
\label{eq:pmom-change-prob-4}
\end{equation}
The integral on the right is exactly equal to
$\LD\Sliver\pi\sliver\exp(-|Y|/\LD)$,
as will be shown shortly.  Hence
$\beta=-(\Gamma/\wspeed_0\sliver\LD)\exp(-|Y|/\LD)$. \
Remembering that
$C=\half\wspeed_0$,
we see that there is an end-to-end difference in \psm\ fluxes
representing a rate of import
$-\half\wspeed_0\sliver\pmom_2 \big|_{x\rightarrow\infty}
=
-\half\wspeed_0\sliver\beta\sliver\pmom_1
=
+(\Gamma\pmom_1/2\LD)\exp(-|Y|/\LD)$\,
of $y$-\psm\ per unit $y$-distance, correct to \oaae.
Recalling the definition of $S$ in
(\ref{eq:mass-flow-prob-1}), with vertical averaging understood,
we sum over all the rays to
find the total recoil force in the $y$~direction as \

\vspace{-0.75cm}

\begin{equation}
\bR
~=~
(\Gamma S/2\LD)
\exp(-|Y|/\LD)\sliver(+\byvec)
~,
\label{eq:recoil-from-qg-rays}
\end{equation}
in agreement with (\ref{eq:force-prob-3}).

The integral on the right of
(\ref{eq:pmom-change-prob-4})
is equal to $\LD$ times the value at $y'=Y/\LD$
of the function
$I(y')$ defined by
$I(y')=\int_{-\infty}^\infty K_0(\Radius)\sliver dx'$
where $x'=x/\LD$ and $y'=y/\LD$ so that
$\Radius^2=x'^2+y'^2$.  Now $K_0(\Radius)$ is equal to its
Laplacian in the
$x',y'$ plane,
except at the origin where the Laplacian has
a delta function $-2\pi\sliver\delta(x')\delta(y')$
in place of the integrable logarithmic singularity in $K_0$ itself.
For any $y'\ne0$ we therefore have

\vspace{-0.25cm}

\begin{equation}
I(y')
~=~
\int_{-\infty}^\infty
\left(
\frac{\partial^2}{\partial x'^2}
+
\frac{\partial^2}{\partial y'^2}
\right)
K_0(\Radius)\sliver dx'
~=~
\frac{d^2}{dy'^2}
\int_{-\infty}^\infty
K_0(\Radius)\sliver dx'
~=~
\frac{d^2}{dy'^2}I(y')
\label{eq:laplacian-trick}
\end{equation}
and, taking the delta function into account, we have
for all $y'$ from $-\infty$ to $+\infty$

\vspace{-0.25cm}

\begin{equation}
\frac{d^2}{dy'^2}I(y')
-
I(y')
~=\;
-2\pi\delta(y')
~,
\label{eq:laplacian-trick-ode}
\end{equation}
whose solution evanescent at infinity is
$I(y')=\pi\exp(-|y'|)$,
corresponding to the result asserted.

\vspace{-0.5cm}

\section{Concluding remarks}
\label{sec:conclu}

Despite their restricted parameter range,
the problems studied here are enough to remind us
that remote recoil, as such, is
generic and ubiquitous.
Remote recoil
will occur whenever
wave-induced mean flows extend outside wavetrains or wave packets
and advect coherent vortices.
Remote recoil is excluded, or made subdominant, by
asymptotic theories of wave--current interactions that assume
slowly-varying mean currents
with a single large length-scale and correspondingly weak
vorticity or PV anomalies.

The main question left open by this work concerns the
scope of the \psm\ rule.
As remarked at the end of \S\ref{sec:imp-psm},
the rule is known to be valid in a wider range of cases than those
considered here,
even though in a still wider context
there are known
exceptions
including the
case of one-dimensional sound waves in a rigid tube,
as noted long ago in
Brillouin's classic works on radiation stress.
Further exceptions include
the internal-gravity-wave problem studied in \citet{McIntyre:1973}
and the rotating problems studied in
\citet[\& refs.]{Thomas:2018},
in some ways similar to our problem~(iii).
The failure of the rule in these latter cases,
and in Brillouin's, is related to \oaa\ mean
pressure reactions from
confining boundaries (more detail
in Appendix~\ref{sec:appc} below).  We may similarly expect failure of the rule in
laboratory experiments such as those of \citet{Humbert:2017},
conducted in tanks or channels with confining walls that can support \oaa\
mean pressures.
Section~\ref{sec:imp-psm} reminds us that
the impulse--\psm\ theorem depends
on having a sufficiently large fluid domain
enclosing the regions occupied by waves and vortices.

For the reasons indicated at the end of
section~\ref{sec:imp-psm}, even in a large domain
the scope
of the \psm\ rule in wave--vortex interactions
is very much a nontrivial question calling for further research,
probably involving numerical experimentation along the lines of
the strong-refraction experiments of \citet{Coste:1999}.
Even though the Kelvin impulse concept depends on
banishing large-scale pressure-field
adjustments to infinity,
the basic \te\ associated with the Magnus relation
(\ref{eq:magnus-force}),
that
of applying a force to move a vortex core,
works, by contrast, in a relatively local way.
This poses not only a technical but also a nontrivial conceptual challenge.

Regarding quantum vortices, it would be interesting to see
how the present analysis of problem~(ii) extends to the
Gross--Pitaevskii superfluid model, a context in which problem~(i)
was studied in \citet{Guo:2014}.
In the corresponding version of problem~(ii) we can expect to find the same
\nil\
and the same caveats regarding the \ab\ effect, pointing to a
remote-recoil contribution
in addition to the Iordanskii force.
The Gross--Pitaevskii model
provides a simple representation of
quantum vortex cores \citep{Berloff:2004},
whose supersonic flow velocities
might vitiate any attempt at a weak-refraction theory,
even though the small core size might, on the other hand,
imply that bodily advection of the core --
back and forth by a larger-scale wavemotion as well as
persistently by the mean flow --
could still be a useful simplifying feature.

\vspace{0.2cm}

{\small
\noindent\textit{Acknowledgements:}
Pavel Berloff provided the first stimulus to embark on this study.
I thank him and
Natalia Berloff,
Oliver B\"uhler,
Victor Kopiev,
Hayder Salman,
Mike Stone,
Jim Thomas,
Jacques Vanneste
and
Bill Young
for their interest
and for their very useful comments -- some of them on
substantial technical points -- during the writing and revision of the paper.
I should also like to thank the three
referees, whose comments were extremely challenging and
of great value in helping me to sharpen the presentation,
in the course of two major revisions.
}

\vspace{-0.6cm}

\appendix

\section{The Schr\"odinger equation and the phase function (\ref{eq:disloc-wavefield})}
\label{sec:appa}

In the quantum problem originally studied by
\citet{Aharonov:1959}, the wave~field
$\phi=\exp(\imag\sliver\Phi)$ with $\Phi$ defined by
(\ref{eq:disloc-wavefield})
is not only a far field but also an exact solution.
For reasons that are obvious from figure~\ref{fig:wavecrests-exaggerated},
the quantum literature often calls it a ``dislocated'' wave~field.
In the quantum problem there is no restriction to
small $\alpha$.  That is easily verified;
the relevant Schr\"odinger equation  can be written in suitable units as

\vspace{-0.3cm}

\begin{equation}
\imag \,\frac{\partial\phi}{\partial t}
~+~
\frac{\wspeed_0}{\, k_0}
\Antisliver
\left(
   \bnabla
   +
   \imag \alpha \radius^{-1}\bthetavec
\right)^{\Antisliver\raisebox{-2pt}{\scriptsize 2}}
\!\phi
~=~
0
~,
\label{eq:schro-eqn}
\end{equation}
where the square denotes a scalar product.
When $\phi=\exp(\imag\sliver\Phi)$,
with the error term deleted from (\ref{eq:disloc-wavefield}), we have \
$\partial\phi/\partial t = -\imag \wspeed_0 k_0\phi$ \
and
$\nabla\phi = (\imag k_0\bxvec - \imag \alpha \radius^{-1}\bthetavec)\phi$, \
satisfying (\ref{eq:schro-eqn}) exactly.
This wavefunction $\phi$ is part of a solution to
(\ref{eq:schro-eqn})
that describes nonrelativistic electrons
going past an infinitely long, thin magnetic solenoid,
whose total magnetic flux and magnetic vector potential
$\propto\radius^{-1}\bthetavec$
play the roles of
$\Gamma$ and $\bu_0$ in the vortex problem.

\begin{figure}
\centering
\vspace{0.1cm}
\includegraphics[width=6 cm, viewport=90 360 650 750,
clip]{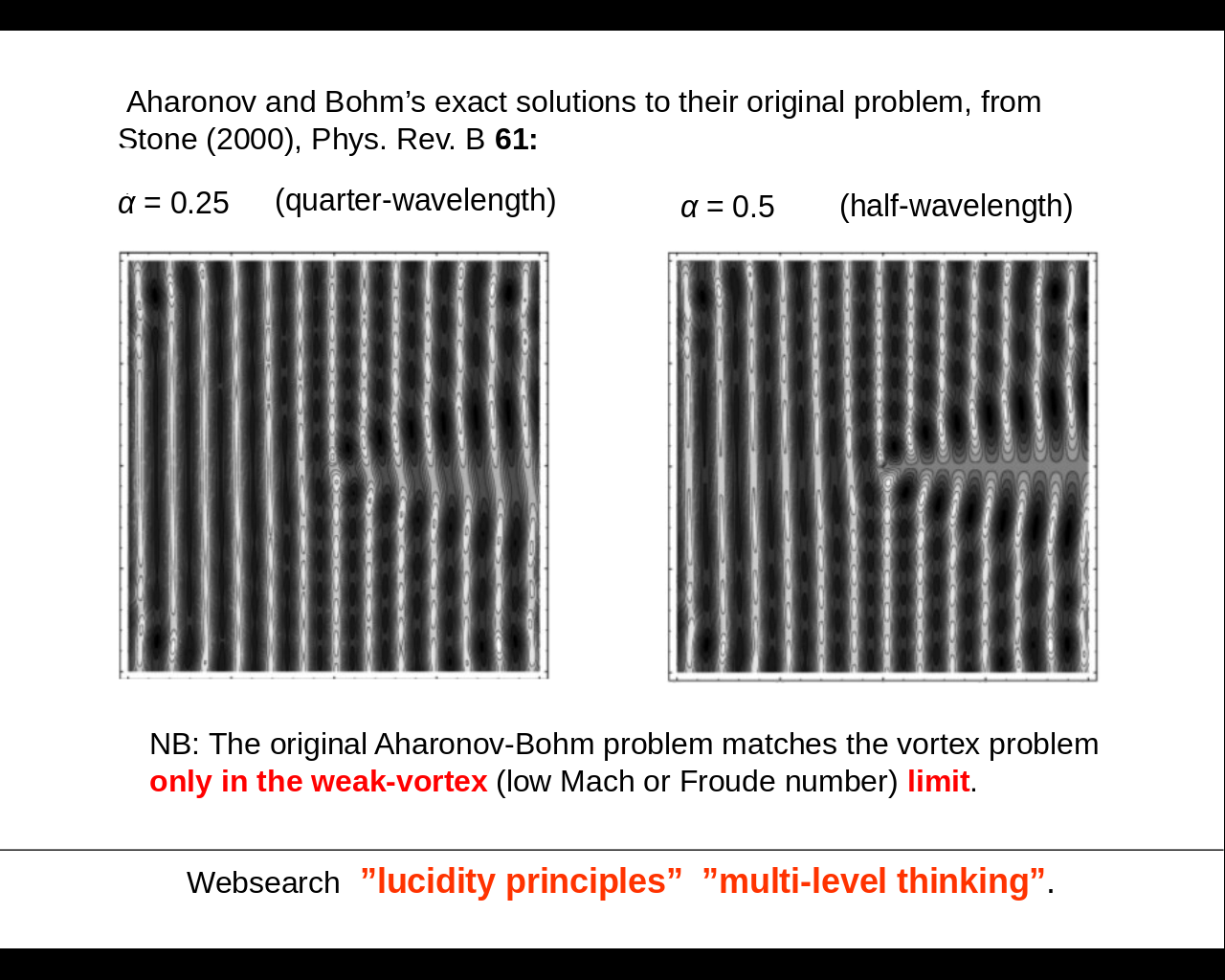}
\caption{Numerical solution of the original \ab\ problem
(\ref{eq:schro-eqn}), from
\citet{Stone:2000a}.  The real part of $\phi$ is plotted.
Here $\alpha=0.25$, just large enough to
make the phase change across the Fresnel wake easily visible.
Also visible, very faintly, is a Born-scattered contribution
recognizable by its approximately circular wavecrests.
Reprinted, with permission, from figure~1 of \citet{Stone:2000a};
copyright 2000 by the American Physical Society.
}
\vspace{-0.5cm}
\label{fig:stone-ab-pic}.
\end{figure}

In the complete solution, originally derived by
Aharonov \& Bohm and generalized to a
solenoid of arbitrary diameter by \citet{Berry:1980},
there is in addition a Fresnel diffractive wake and a smaller,
$O(\radius^{-1/2})$
contribution outside the wake region. 
The Fresnel wake is exactly centred on the positive $x$~axis,
for arbitrary $\alpha$,
and smooths out the discontinuity in $\Phi$.
In the thin-solenoid case the smaller, $O(\radius^{-1/2})$ contribution
is describable as Born scattering off the solenoid, whose approximately
circular wavecrests are faintly visible in figure~\ref{fig:stone-ab-pic},
from a paper by \citet{Stone:2000a}.  The figure shows
a numerical calculation of the thin-solenoid solution of (\ref{eq:schro-eqn})
with $\alpha=0.25$, large enough to make
visible the phase change across the Fresnel wake.

All these features are qualitatively the same as
those found by FLS in their analysis for $k_0\sliver\radius_0 \ll 1$
of the linear wave field in the vortex problem.
However, in stark contrast with the Schr\"odinger problem,
the wave~field becomes qualitatively different \citep[e.g.][]{Coste:1999}
as soon as $\alpha$ goes outside the  very restricted range of values
permitted by (\ref{eq:alpha-super-small}), when $\frma \ll 1$
as well as $k_0\sliver\radius_0 \ll 1$.

\vspace{-0.55cm}

\section{Secular changes in problem~(iii)}
\label{sec:appb}

In deriving (\ref{eq:for-prob-3}) in \S\ref{sec:prob3} we ignored a
subtlety worth remarking on.
The argument for taking $\qtilde = q_0$, even though correct
within the \qg\ framework,
does not by itself exclude secular changes in $\qtilde$
over very long times in the \emph{exact} dynamics of
problem~(iii).  However, such changes can be excluded by
appealing to the exact conservation of the Kelvin circulation
around material loops of all sizes, shapes and orientations as
fluid particles travel around the vortex, and in and out of the
wavetrain.  The \oaaaa\ terms neglected in going from
(\ref{eq:3d-exact-mean-vort-cdot-form-vert-cpt-divh}) to
(\ref{eq:3d-mean-vort-cdot-form-vert-cpt-divh-integd})
describe only slight, reversible distortions, within the
wave layer, of such material loops and of the absolute vortex
lines threading them -- the inertially-stiff lines of
$\bfcoriolis + \bomegatilde$. \ Equation (\ref{eq:pv-cons})
is only an approximate expression of the exact statement that the
Kelvin circulation is constant for all material loops,
including those whose parts outside the wavetrain
lie in horizontal planes, at all altitudes $z$.
The circulation of such loops cannot change
secularly unless the qualitative geometry of the picture changes,
such that the loops and the
absolute vortex lines deform irreversibly.
Physically, this would correspond to the presence of
large-amplitude \emph{breaking} waves,
in this case breaking surface gravity waves, or
breaking inertia waves, or both.  As shown by GLM theory,
the irreversible deformation of otherwise wavy material contours
can usefully be taken as the defining property of wave breaking
\citep{McIntyre:1985}.
Our \te s assume that no such wave breaking occurs.

\vspace{-0.55cm}

\section{Asymptotic validity of equation~(\ref{eq:for-prob-3})}
\label{sec:appc}

As well as using $\Ro \ll 1$ in going from
(\ref{eq:3d-exact-mean-vort-cdot-form-vert-cpt-divh}) to
(\ref{eq:3d-mean-vort-cdot-form-vert-cpt-divh-integd-expanded}),
the derivation of (\ref{eq:for-prob-3}) used
overall hydrostatic balance to determine
horizontal pressure gradients on the bottom
as $\rho\sliver g\bnablah\htilde$,
together with
geostrophic balance to give (\ref{eq:streamfunctiondef}) with
$\psitilde = g(\htilde-H)/\fcoriolis$
beneath the wavetrain and elsewhere.
The Taylor--Proudman effect
extends this picture upward into the wavetrain via the
stiffness of the
vortex lines of $\bomegatilde + \bfcoriolis$,
which bend away from the vertical
only slightly, through small angles $O\sliver(\Ro)$. \

Before proceeding to the asymptotic justification of
(\ref{eq:streamfunctiondef}) and (\ref{eq:for-prob-3}),
we note that overall hydrostatic balance does actually fail,
along with the impulse--\psm\ theorem and the \psm\ rule,
in the somewhat similar
problems studied in \citet[\& refs.]{Thomas:2018}.
Those problems assume rotating shallow~water dynamics
for the wavemotion as well as for the mean flow.
The failure is
due to the confinement of the wavetrain
by the lower boundary.  Other cases
of confinement
by boundaries and consequent pseudomomentum-rule failure include
the classic case of one-dimensional sound waves in a rigid tube,
with a wavemaker at one end and an absorber at the other
\citep[e.g.][B14 \S12.2.2]{Brillouin:1936,McIntyre:1981}.
In
problems like that of Thomas \etal,
the lower boundary
gives the radiation-stress field a
foothold -- a bottom boundary
to react against -- allowing the
stress divergence to push or pull vertically
on the complete fluid column
and to disrupt overall hydrostatic balance so as to change the \oaa\
pressure gradients on the bottom.
This in turn produces
additional terms on the right of equations like (\ref{eq:for-prob-3})
governing potential-vorticity inversion,
breaking the impulse--\psm\ theorem by breaking the connection
between
$\psitilde$ and $g(\htilde-H)/\fcoriolis$.
Remote recoil is still generic, however.
Here we are using the term ``radiation stress''
in the slightly loose sense of any
wave-induced momentum flux that arises from averaging the
equations of motion in some way,
rather than in the stricter sense
adopted in Brillouin's writings and for instance in
\citet{LonguetHiggins:1964b}, in
B14 \S10.5, and in 
\citet[][hereafter AM78, \S8.4]{Andrews:1978a},
to mean the sole effect of the waves on the mean flow --
which is definable in some but not all
wave--mean interaction problems.

In problem~(iii),
the impulse--\psm\ theorem does hold
and with it the \psm\ rule --
as was independently confirmed in \S\ref{sec:prob3} --
essentially because
the foothold effect is
too small to disrupt overall hydrostatic balance,
thanks to sufficient separation between the lower boundary
and the wavetrain such that $\exp(-kH)$ can be neglected.
To verify this in detail
and to check for other possible errors
it is simplest, again, to work within the GLM framework,
thereby avoiding the complications that come
from the intersection of the free surface with the horizontal
Eulerian coordinate surface $z=0$, which we take as the
undisturbed free surface.
We would like to demonstrate asymptotic validity
not only for problem~(iii), but also for the
wider variety of wave--vortex configurations
covered by the impulse--\psm\ theorem in \S\ref{sec:imp-psm}.
For scale-analytic purposes we use
$\ktyp$ and $\ktyp\wspeed$ to denote a typical wavenumber and frequency,
whose orders of magnitude are unaffected by weak refraction.

Clearly
$\amp$, $\frma$, $\Ro$, $\fcoriolis/\ktyp \wspeed$ and $\exp(-\ktyp H)$
must all be treated as small parameters, the last two in order to
use deep-water wave dynamics with Coriolis effects neglected and to
guarantee
negligibility of the foothold effect.
We would like to let all five parameters
tend toward zero, keeping $\amp\ll\frma$,
for a given geometry of the vortex core, or cores,
and the incident wave~field.

For simplicity's sake we
restrict attention to cases in which

\vspace{-0.5cm}

\begin{equation}
\amp
~\ll~
\Ro
~\sim~
\fcoriolis/\ktyp\wspeed
~\sim~
\frma
\:\ll\; 1
\label{eq:asy-regime-appc}
\end{equation}
in the limit.  It will prove expedient, however, to allow
$\exp(-\ktyp H)\,\lesssim\,\frma$.\,
As in (\ref{eq:rossby-number}) we take

\vspace{-0.95cm}

\begin{equation}
\Ro
~=~
U/\antisliver\fcoriolis\radius_0
~.
\label{eq:rossby-number-frma-appc}
\end{equation}
Given geometry means that horizontal scales such as $\width$ and
$\radius_0$ will be held fixed in the limit.
We therefore need not distinguish among those scales, and will take
$\radius_0$ as their representative.
It is convenient also to fix $\fcoriolis$ and $\wspeed$, and to take
$U$ toward zero like $\frma$.
Then (\ref{eq:asy-regime-appc}) implies that
we must take
$\ktyp$
toward infinity
like $\frma^{-1}$.
The meaning of ``given incident wave~field''
will therefore have to be relaxed to
mean a given amplitude distribution while $\ktyp \rightarrow\infty$,
consistent with ray theory.
We must also take gravity $g$
toward infinity like $\frma^{-1}$, because $\wspeed^2=g/\ktyp$.
Restated in a dimensionally consistent way, these conditions
can be summarized as

\vspace{-0.45cm}

\begin{equation}
U \,\sim\,
\wspeed\Sliver\frma
\,\sim\,
\fcoriolis\radius_0\Sliver\frma
,~~~~
\ktyp \,\sim\,
\radius_0^{-1}\frma^{-1}
,~~~~
g \,\sim\,
\wspeed^2\ktyp
 \,\sim\,
\wspeed^2\sliver\radius_0^{-1}\frma^{-1}
\label{eq:diml-condns-appc}
\end{equation}
as $\frma \rightarrow 0$.
The assumption \,$\exp(-kH)\lesssim\frma$\, implies that
$\ktyp H\gtrsim|\antisliver\ln\sliver\frma|$
and hence that
$H\Antisliver\gtrsim\radius_0\Sliver\frma\sliver|\antisliver\ln\frma|$ and
$\LD=(gH)^{1/2}/\fcoriolis\gtrsim\radius_0\sliver|\antisliver\ln\frma|^{1/2}$\Antisliver,
which allows enough flexibility to accommodate our illustrative results
(\ref{eq:core-advec-prob-3})--(\ref{eq:recoil-from-qg-rays})
alongside the more general wave--vortex configurations
considered in \S\ref{sec:imp-psm}.
It is convenient also to assume that $H\lesssim\radius_0$,
though this is hardly a significant restriction since
$H\sim\radius_0$ would
correspond to $\LD\sim\radius_0\sliver\frma^{-1/2}$,
greatly exceeding any other horizontal scale.

We assume that the
pressure on the free surface is constant, with or without
disturbances, and take the constant
to be zero without loss of generality.
Within the wavetrain there is a
three-dimensional \oaaeo\ radiation stress or wave-induced momentum
flux $\Pi_{ij}$, say,
which dies off exponentially with depth
like $\exp(2kz)$,
for deep-water waves with vertical structure $\exp(kz)$,
as well as vanishing at the free surface $z=0$.
The Cartesian-tensor indices
$i,j$ now take values $(1,2,3)$, corresponding to $(x,y,z)$.
The sign convention will be such that the force per unit volume felt by the
mean flow is $-\sliver\Pi_{ij,j}$.
The most convenient formula
for $\Pi_{ij}$\sliver,
which is an \oaa\ wave property, is

\vspace{-0.2cm}

\begin{equation}
\Pi_{ij}
~=~
-\pbarl\antisliver
\big\{
\half\overline{(\xi_\lind\xi_\mind)}
_{\raisebox{1.3pt}{\scriptsize\antisliver$,\lind\mind$}}
\sliver\delta_{ij}
-
\overline{(\xi_{\lind,i}\xi_{j})}
_{\raisebox{1.3pt}{\scriptsize\antisliver$,\lind$}}
\big\}
-
\overline{p^\ell\xi_{j,i}}
\label{eq:rad-stress-def}
\end{equation}
where $\pbarl$ is the
Lagrangian-mean pressure and $p^\ell$ the
\oa\ Lagrangian disturbance pressure,
while $\bxi$ is the \oa\ disturbance particle-displacement field, with
Cartesian components \ $\bxi=(\xi_1,\,\xi_2,\,\xi_3)$ \
and zero divergence
$\xi_{\lind,\lind}=0$ correct to \oa. \
The formula (\ref{eq:rad-stress-def}) can be read off
from AM78 (8.6), (8.10) and (9.3), or from
B14 (10.43), (10.57), (10.73), (10.77) and (10.84).\footnote{In
AM78, $\Pi_{ij}$ is denoted by~$-R_{ij}$,
and in B14 by $\tilde\Pi_{ij}-\pbarl\sliver\delta_{ij}$.
When using AM78 (8.10) we can neglect the divergence of $\bxi$
as well as an \oaa\ term $k_{ij}$,\,
before substituting into (8.6) and discarding
terms $\propto\amp^3$ or higher.
In B14, (10.73) is rewritten as
$K_{km}=J\delta_{km} -\xi_{i,k}K_{im}$ before
substituting it into
(10.84)
in the same way.  Then use 
is made of (10.43), (10.57), and~(10.77).
The equation numbers in B14 correspond to
   (10.43), (10.57), (10.71), (10.75) and (10.82)
in the original, 2009 edition.
Though not needed here,
it may be of interest to note that
substitution of
the leading order deep-water plane wave structure
(which has $p^\ell=0$)
into the horizontal components of
(\ref{eq:rad-stress-def})
leads to
the standard \oaa\ result
$\int\Antisliver\bpmom\Sliver\bC\sliver dz$ for the
depth-integrated horizontal momentum flux
\citep[e.g.\ (24),\,(33) of][]{LonguetHiggins:1964b}.

}

We use a ray-theoretic description of the waves,
relative to suitably-oriented horizontal axes.
The $x$ or $x_1$ axis is chosen parallel to the local wavenumber,
whose magnitude
is asymptotically large like $\frma^{-1}$
according to (\ref{eq:diml-condns-appc}). \
Zooming in to the local plane-wave structure, we have

\vspace{-0.4cm}

\begin{equation}
(\xi_1,\,\xi_2,\,\xi_3)
\,=\,
b \exp(kz)\sliver \{\cos\Phi + O(\frma), \
O(\frma), \
\sin\Phi + O(\frma)\}
\label{eq:wave-structure-appc}
\end{equation}
where $\Phi=k(x - \wspeed t) \ + $ const.,
with $k$, $\wspeed$ and the displacement amplitude $b$ all locally constant.
We take
$\amp = bk$,
so that $\amp\ll\frma$ is the dimensionless wave slope.
The relative errors \oeps\
include weak-refractive effects as well as
a small transverse displacement $\xi_2 = O(\frma)$ whose magnitude
arises from
our assumption in (\ref{eq:asy-regime-appc})
that $\fcoriolis/\ktyp \wspeed \,\sim\, \frma$, in agreement
with Pollard's exact solution, which incidentally has
$p^\ell$ exactly zero.  However, to allow for weak refraction
we will use a more conservative estimate
$p^\ell\lesssim \frma \rho gb\sliver\exp(kz)$,
which is \oeps\ times the Eulerian disturbance pressure.

The overbars in (\ref{eq:rad-stress-def}) are to be read as
Eulerian phase averages over the local wave structure,
in the standard way.
Notice that if $\pbarl$ were constant and $p^\ell$ zero
then the divergence $\Pi_{ij,j}$ would vanish.  Therefore an additive constant
in the pressure has no effect on
the dynamics, confirming that, without loss of generality,
we may take $\Pi_{ij}=0$ at the free surface.
Neglecting \oaa\ contributions to $\pbarl$, we can replace it by
$-\rho gz$ so that correct to \oaa\

\vspace{-0.4cm}

\begin{equation}
\Pi_{ij}
~=~
\rho gz\sliver
\big\{
\half\overline{(\xi_\lind\xi_\mind)}
_{\raisebox{1.3pt}{\scriptsize\antisliver$,\lind\mind$}}
\sliver\delta_{ij}
-
\overline{(\xi_{\lind,i}\xi_{j})}
_{\raisebox{1.3pt}{\scriptsize\antisliver$,\lind$}}
\big\}
-
\overline{p^\ell\xi_{j,i}}
~.
\label{eq:rad-stress-def-approx}
\end{equation}

\noindent
The resultant vertical force on a complete fluid column
per unit horizontal area is

\vspace{-0.3cm}

\begin{eqnarray}
\int_{-H}^{~\:0} \Pi_{3j,j}\Sliver dz
~=\,
\int_{-H}^{~\:0} 
\big\{
  [\half \rho gz\overline{(\xi_\lind\xi_\mind)}
  _{\raisebox{1.3pt}{\scriptsize\antisliver$,\lind\mind$}}
  ]_{\raisebox{-1pt}{\scriptsize\antisliver$,3$}}
  -
  [\rho gz\overline{(\xi_{\lind,3}\xi_{j})}
  _{\raisebox{1.3pt}{\scriptsize\antisliver$,\lind$}}
  ]_{\raisebox{-0.5pt}{\scriptsize\antisliver$,j$}}
  -
  [\overline{p^\ell\xi_{j,3}}
  ]_{\raisebox{-0.5pt}{\scriptsize\antisliver$,j$}}
\big\}
\sliver dz
\nonumber\\
~=~\,
\big\{
  \half\rho gH\overline{(\xi_\lind\xi_\mind)}
  _{\raisebox{1.3pt}{\scriptsize\antisliver$,\lind\mind$}}
  -
  \rho gH\overline{(\xi_{\lind,3}\xi_{3})}
  _{\raisebox{1.3pt}{\scriptsize\antisliver$,\lind$}}
  -
  \overline{p^\ell\xi_{3,3}}
\big\}
\big|_{z=-H}
  \hspace{1cm}  \phantom{|}
\nonumber\\
\;-
\int_{-H}^{~\:0}
\big\{ \rho gz
  \overline{(\xi_{\lind,3}\xi_{\gamma})}
  _{\raisebox{1.3pt}{\scriptsize\antisliver$,\lind\gamma$}}
  +
  [\overline{p^\ell\xi_{\gamma,3}}
  ]_{\raisebox{-0.5pt}{\scriptsize\antisliver$,\gamma$}}
\big\}
\Sliver dz
  \hspace{1.2cm}  \phantom{|}
\label{eq:vert-force}
\end{eqnarray}
where the greek index $\gamma$ runs from 1 to 2,
but $j$, $\lind$ and $\mind$ still from 1 to 3.
This expression
is more convenient than the alternative expression
obtainable by applying derivatives
to the factor $\rho  gz$ only, in the first line, giving a result that
looks simpler but obscures the foothold effect,
the expression
in the second line.

Vertical derivatives
$\partial/\partial z = \partial/\partial x_3$
have order of magnitude $\sim k\sim\frma^{-1}\Antisliver$,\Sliver\
and horizontal derivatives
order unity or less,\, $\lesssim\frma^0$,\,
as $\frma\rightarrow 0$,\, because horizontal
scales such as $\radius_0$ and $\width$ are being held fixed, while
$\LD\gtrsim\radius_0\sliver|\antisliver\ln\frma|^{1/2}\Antisliver$.\,
In the last term of the foothold contribution on the second line
of (\ref{eq:vert-force}) we use our conservative estimate
$p^\ell\lesssim \frma \rho gb\sliver\exp(kz)$,
and the assumptions
$\ktyp\sim\radius_0^{-1}\frma^{-1}$ and
$\exp(-kH)\lesssim\frma$ made in
(\ref{eq:asy-regime-appc})--(\ref{eq:diml-condns-appc}),
to show that the term in question has magnitude
$\lesssim \frma\sliver \rho g\sliver b^2 k \exp(-2kH)
 \lesssim \frma^2\sliver \rho g\sliver b^2\antisliver/\radius_0$.
The first two terms combine to give a larger estimated magnitude
$\lesssim \frma\sliver \rho g\sliver b^2\antisliver/\radius_0$, as shown next.

In the first two terms we note
that the largest, vertical-derivative contributions
$
\half\rho gH\overline{(\xi_3\xi_3)}
_{\raisebox{1.3pt}{\scriptsize\antisliver$,33$}}
$
and
$
-\rho gH\overline{(\xi_{3,3}\xi_{3})}
_{\raisebox{1.3pt}{\scriptsize\antisliver$,3$}}
$
cancel each other
to leading order.  This is because of~the
special structure of deep-water waves
and would not be the case in, for instance,
the problems studied by Thomas \etal\, \
For the local plane-wave structure we have
$\raisebox{-1pt}{$\overline{\sin^2\Phi}$} \,=\,
 \raisebox{-1pt}{$\overline{\cos^2\Phi}$} \,=\, \half$,
hence $\half\overline{(\xi_3\xi_3)}=\quarter b^2\exp(2kz)$,
with relative error \oeps.
The vertical second derivative 
$\half\overline{(\xi_3\xi_3)}
_{\raisebox{1.3pt}{\scriptsize\antisliver$,33$}}
= \quarter b^2\sliver 4k^2\exp(2kz)
= b^2k^2\exp(2kz)
= \overline{(\xi_{3,3}\xi_{3})}
_{\raisebox{1.3pt}{\scriptsize\antisliver$,3$}}
$\Sliver.\,
Therefore the sum of the first two terms
has its order of magnitude reduced
by a factor $\frma$ or less, and can be estimated as
$\lesssim \frma\sliver \rho g\sliver b^2 k^2H \exp(-2kH)$.
Using our assumptions
$\ktyp\sim\radius_0^{-1}\frma^{-1}$, $\exp(-kH)\lesssim\frma$, and
$H\lesssim\Sliver\radius_0$,
we have
$\frma\sliver \rho g\sliver b^2 k^2H \exp(-2kH)
 \lesssim \frma\sliver \rho g\sliver b^2\antisliver/\radius_0$
as asserted.  Thus the entire foothold contribution,
the second line of (\ref{eq:vert-force}),
can be estimated as
$\sim \frma\sliver \rho g\sliver b^2\antisliver/\radius_0$
at most.

In the vertically integrated, non-foothold contribution
in the third line of (\ref{eq:vert-force}), each term
has magnitude
$\lesssim \frma\sliver\rho g\sliver b^2\antisliver/\radius_0$ also.
To check this for the term in $p^\ell$, we may take
$\int\!...\: dz \sim \ktyp^{-1}\antisliver$,\,
and as before use
$p^\ell\lesssim\frma\sliver\rho g\sliver b\sliver\exp(kz)\sim
\frma\sliver\rho g\sliver b$
in the integrand, so that
$[\overline{p^\ell\xi_{\gamma,3}}
 ]_{\raisebox{-0.5pt}{\scriptsize\antisliver$,\gamma$}}
\lesssim \frma \rho gb^2\ktyp\antisliver/\radius_0
$,
the factors $\ktyp$ and $\radius_0^{-1}$ coming from
the vertical and horizontal derivatives respectively.
Integration removes the factor $\ktyp$, leaving a contribution
$\lesssim \frma\sliver \rho g\sliver b^2\antisliver/\radius_0$.

In the other term, the first term on the third line, we have
$\overline{(\xi_{\lind,3}\xi_{\gamma})}
 _{\raisebox{1.3pt}{\scriptsize\antisliver$,\lind\gamma$}}
\lesssim \frma b^2 \ktyp^2\antisliver/\radius_0\sliver
$,
with a factor $\ktyp^2$
since among the three derivatives at most two are vertical,
as happens in the contribution with $\lind=3$.
The factor $\frma$
comes from
the \oeps\ relative magnitude of $\xi_\gamma$ when $\gamma=2$
or, when $\gamma=1$, from
the phase difference between $\xi_{3,3}$ and $\xi_1$,
which is $\pi/2 + O(\frma)$.\,
So averaging their product produces a factor $\frma$.
With $\int\!...\: dz \sim \ktyp^{-1}\antisliver$,\,
and $\rho gz \sim \rho g\ktyp^{-1}$,
this term and therefore the whole third line
$\:\lesssim \frma\sliver \rho g\sliver b^2\antisliver/\radius_0\,$
as asserted.

In summary, then,
the resultant vertical force (\ref{eq:vert-force}) per unit horizontal area
$\lesssim \frma\sliver \rho g\sliver b^2\antisliver/\radius_0$.
This is the greatest amount by which
the pressure on the bottom boundary can depart from its hydrostatic
value $\rho g\htilde$. \ Let $\errorpsi$ be the corresponding error in
$\psitilde = g(\htilde-H)/\fcoriolis$; then $\errorpsi\lesssim
\frma\sliver g\sliver b^2\antisliver/(\fcoriolis\radius_0)$. \
In the operator $(\nablahsq - \LD^{-2})$
on the \lhs\ of (\ref{eq:for-prob-3})
  the relevant horizontal scales
  are either fixed $\sim\radius_0$, in the limit $\frma\rightarrow0$,
  or expand slightly because
  $\LD\gtrsim \radius_0 |\antisliver\ln\frma|^{1/2}\antisliver$.\,
  So the error
  on the \lhs\ of (\ref{eq:for-prob-3})
is no greater than $\errorpsi$ divided by
$\radius_0^2$ as $\frma\rightarrow0$; so the error
$\lesssim \frma\sliver g\sliver b^2\antisliver/(\fcoriolis\radius_0^3)$. \
To neglect this error, we need to show that it is small in comparison with
$\bzvec\sliver\cdott\Antisliver\bnabla\cross\langle\bpmom\rangle$ on
the \rhs\ of (\ref{eq:for-prob-3}).
Estimating
$\bzvec\sliver\cdott\Antisliver\bnabla\cross\langle\bpmom\rangle$ as
$\sim\langle\bpmom\rangle_{\rm typ}/\width
 \sim\langle\bpmom\rangle_{\rm typ}/\radius_0$,
where
$\langle\bpmom\rangle_{\rm typ}$ is a typical magnitude of
$\langle\bpmom\rangle$,
we therefore need to show that

\vspace{-0.35cm}

\begin{equation}
\frma\sliver g\sliver b^2\antisliver/(\fcoriolis\radius_0^3)
\;\ll\;
\langle\bpmom\rangle_{\rm typ}/\radius_0
\label{eq:verify-hydrostat}
\end{equation}
as $\frma\rightarrow0$. \ Now
$\langle\bpmom\rangle_{\rm typ}\sim g\sliver b^2\antisliver/(\wspeed H)$
since in ray theory $\langle\bpmom\rangle$ is $\wspeed^{-1}$ times
the wave-energy per unit horizontal area, $\sim \rho g\sliver b^2$,
divided by $\rho H$.  So in (\ref{eq:verify-hydrostat})
the ratio of the \lhs\ to the \rhs\ is
$\frma\sliver\wspeed H/(\fcoriolis\radius_0^2)$; and, recalling that
$\frma\sliver\wspeed \sim U$ and that
$\Ro\sim U/(\fcoriolis\radius_0)$, we see that 
\,$\frma\sliver\wspeed H/(\fcoriolis\radius_0^2)
\,\sim\,\Ro\Sliver H/\radius_0\,\lesssim\,\Ro\,\sim\,\frma$.\,
This estimate is sufficient for our purposes, but
is very conservative because it relies again on the assumption
$H\lesssim\Sliver\radius_0$. \
If we restrict $H$ more tightly, to its marginal order of magnitude 
$H\Antisliver\sim\radius_0\Sliver\frma\sliver|\antisliver\ln\frma|$,
then
(\ref{eq:verify-hydrostat}) is satisfied more strongly, with
ratio $\frma^2\sliver|\antisliver\ln\frma|$
instead of $\frma$. \
Either way, (\ref{eq:streamfunctiondef}) and (\ref{eq:for-prob-3})
have now been validated, as required, as leading-order approximations
on the basis of which
Bretherton flows can be computed correct to \oaaeo\ and thence
recoil forces correct to \oaae.

\smallskip

Although the foregoing is sufficient for our purposes,
the results can of course be checked
directly from the vertical component of the GLM
momentum equation, AM78 (8.7a) or B14 (10.82).
In carrying out that check it needs to be remembered
that the GLM divergence effect
raises the Lagrangian-mean altitudes of the free surface
and other isobaric material surfaces.
To leading order, in the local plane wave, the surfaces are raised by
\oaa\ amounts \
$\half \overline{(\xi_3^2)}
 _{\raisebox{1.3pt}{\scriptsize\antisliver$,3$}}
$, \
as is also necessary to account for the
waves' potential energy
$\half\rho\sliver g\Sliver\overline{\xi_3^2}|_{z=0}$ per unit area
\citep{McIntyre:1988}.
The raising of the free surface is accompanied by
a compensating \oaa\ reduction, \
$\half\rho\sliver\overline{(\xi_3^2)}
  _{\raisebox{1.3pt}{\scriptsize\antisliver$,33$}}
$, \
in the mean density $\rhotilde\sliver$ defined such that
$\rhotilde\Sliver dx\sliver dy\sliver dz$ is the volumetric mass element,
consistent with a negligible change in
the total mass overlying a horizontal area element
of the bottom boundary.

\vspace{-0.5cm}

\section{The \oaae\ \psm\ law}
\label{sec:appd}

The two-dimensional pseudomomentum law (\ref{eq:psm-law})
holds to the order of accuracy required in \S\ref{sec:imp-psm},
namely correct to \oaae\ -- the order of magnitude of the refraction term on
the \rhs\ --
as $\Sliver\amp\Sliver$ and $\Sliver\frma\Sliver$ tend toward zero
with $\amp\ll\frma$
for a given geometry of the vortices and incident wave~field,
whose horizontal scales are held fixed in the limit
as in Appendix~\ref{sec:appc}.

The most secure route to (\ref{eq:psm-law})
is to start with its exact GLM counterparts,
in all three problems,
so that we can see precisely what is neglected.
To save space we refer directly to B14's exact GLM equations
(10.123)--(10.126), which are (10.122)--(10.125)
in the original, 2009 edition.  The second of these equations
defines the exact nonadvective flux of \psm,
the exact counterpart of $\fpmom_{ij} - \pmom_i\ubarl_j$ in
(\ref{eq:psm-law})--(\ref{eq:pmom-flux-gp-vel}) above.
We recall that
the nonadvective flux can be rewritten exactly, within the GLM framework,
as an isotropic term $\propto\delta_{ij}$ plus
the wave-induced flux of \emph{momentum}.
This will be useful when considering problem~(iii),
in which the anisotropic part of
the expression on the right of (\ref{eq:rad-stress-def})
will be made use of.

In the gas dynamical version of problems~(i) and (ii)
the motion is strictly two-dimensional.\;
Equation~(\ref{eq:psm-law})
can be read off straightforwardly from its exact counter\-part,
$\rhotilde$\, times B14~(10.126) (see also (10.47)), with
indices $i,\, j$ etc.\ running from 1 to 2. 
The two-dimensional mean density $\rhotilde$ is the same as our $\htilde$
and can be approximated as a constant,
in its product with the refraction term, the third term on the right.
The fractional error involved is small,
\oaaeo\ $+$ \oaoee,
corresponding to absolute error
\oaaaae\ $+$ \oaaeee,
the first term coming from the hard-spring contribution to the
Brillouin radiation stress noted at the end of \S\ref{sec:eqns},
and the second from the Bernoulli pressure drop
surrounding a vortex core.
In the last term on the right of B14~(10.126),
the gradient
$\rhotilde_{,i} = \htilde_{,i}$ is similarly small,
\oaaeo\ $+$ \oaoee, but is
multiplied by the expression in large curly brackets,
which is \oaaeo\
rather than \oaae.
The product \oaaaaeo\ $+$ \oaaee\ is, however, still negligible against
$\amp^2\frma^1$, the magnitude of the refraction term.
The first term on the right
corresponds to $\pmb{\cal F}$ in
(\ref{eq:psm-law}),
with the
irrotational forcing potential
$\phi$
corresponding to $-\chi'$
in (\ref{eq:lind-phi-eqn}).
The second term on the right
is zero, there being
no rotational forcing or
dissipation.
The flux tensor $\fpmom_{ij}$ in our
(\ref{eq:psm-law})--(\ref{eq:pmom-flux-gp-vel})
is given by B14's (10.125) plus the \oaae\ advective flux
$\rhotilde\Sliver\pmom_i\sliver\ubarl_j$, in which $\rhotilde$
can again be approximated as a constant, with the same relative and absolute
errors as in the refraction term.
Thus (\ref{eq:psm-law}) is established correct to \oaae.

In the shallow~water version of problems~(i) and (ii),
the governing equations are the same as in the gas dynamical version
with the ratio of specific heats set to 2, and no more need be said.

For problem~(iii), we need the vertical average of
$\rhotilde\sliver$ times
the horizontal projection of B14 (10.126)
or, more conveniently, of
(10.123), with zero \rhs\ because there is
no rotational forcing or dissipation.
In the horizontal projection, the free suffix $i$ takes values
$i=1,2$, while the dummy suffixes $j$, $k$ and $m$ run from 1 to 3.
The last term on the left of (10.123)
corresponds to $\pmb{\cal F}$,
while the second-last
term, an elastic-energy term,
is zero because the flow is three-dimensionally incompressible.

In the third-last term on the left,
the density $\rho$ is constant
and the factor 
${\overline{(p/\rho)}^{\rm L}}$
can be taken as $-gz$ with error \oaaeo,
while the factor
$\rhotilde_{\Antisliver,\sliver i}/\rhotilde
= \;$ \oaaeo, with
$\rhotilde$ now the three-dimensional GLM mean density,
which contains an \oaaeo\ contribution
from the GLM divergence effect recalled at the end of Appendix~\ref{sec:appc}.\;
This contribution is significant
in the third-last term only.  Everywhere else it represents a negligible
fractional error in $\rhotilde\Sliver$.\,
The third-last term can be simplified to
$\rho^{-1}$ times
$-gz\sliver\rhotilde_{,i}~+$ \oaaaaeo, whose vertical
average is the horizontal gradient of
$-\sliver\langle\sliver gz\sliver(\rhotilde\sliver-\rho) \rangle$,
with error \oaaaaeo.\;
This gradient can be incorporated without further error
into the flux divergence $\,\bnablah\cdott\bfpmom\,$ of our (\ref{eq:psm-law})
(in which vertical averaging is understood),
via a horizontally isotropic contribution
$-\sliver\langle\sliver gz\sliver(\rhotilde\sliver-\rho) \rangle\sliver\delta_{ij}$
to the averaged flux itself,
$\fpmom_{ij}$.

The second term on the left of
B14~(10.123), a wave kinetic energy term,
can be treated in the same way, giving another isotropic contribution
to the vertically-averaged flux $\fpmom_{ij}$ in (\ref{eq:psm-law}).
In the advection term
$\bubarl\cdott\Antisliver\bnabla\sliver\pmom_i
\,=\,\sliver 
\bnabla\cdott\Antisliver(\bu_0\sliver\pmom_i) \ + $ \oaaaaeo, \
the $z$-independent factor $\bu_0$ can be taken outside
the vertical average, as can also be done in the refraction term 
$\ubarl_{k,i}\pmom_k = u_{0\sliver \raisebox{-1.5pt}{\scriptsize$k,i$}}\pmom_k \ +$ \oaaaaeo.\;
Finally, we note that the $\partial/\partial z$
contribution to the three-dimensional flux divergence,
$\rhotilde\sliver$ times
the fourth term on the left of (10.123),
has vertical average zero because of our assumptions,
spelt out in Appendix~\ref{sec:appc}, that
$\exp(-2kH)$ is negligible and that
the pressure vanishes or is constant at the free surface,
so that the nonadvective $13$ and $23$ components
of the three-dimensional \psm\ flux
defined in BM~(10.124) vanish there.
As already mentioned, these anisotropic components are equal to the
corresponding components of the
wave-induced flux of momentum, which
correct to \oaa\ are given
by the anisotropic terms in (\ref{eq:rad-stress-def})
or (\ref{eq:rad-stress-def-approx}).

The foregoing is enough to establish for problem~(iii)
that our (\ref{eq:psm-law}), with vertical averaging understood,
holds to the order of accuracy required in \S\ref{sec:imp-psm}.
However, it may be of interest to note that, to leading order
under the scaling assumptions of Appendix~\ref{sec:appc},
the quantity
$\sliver\langle gz(\rhotilde\sliver-\rho) \rangle$
is $H^{-1}$ times the
potential energy of the deep-water waves per unit area,
replacing the elastic energy in the gas dynamical system
and, in the isotropic part of $\fpmom_{ij}$,
cancelling the wave kinetic energy to leading order
as expected from averaged-Lagrangian considerations.
Using $\rhotilde = \rho\Sliver(1 -\half \overline{(\xi_3^2)}
 _{\raisebox{1.3pt}{\scriptsize\antisliver$,33$}})
$,
following on from the end of Appendix~\ref{sec:appc},
and continuing to neglect $\exp(-2kH)$
we have, using integration by parts,

\vspace{-0.4cm}

\begin{equation}
H \langle \sliver gz(\rhotilde\sliver-\rho) \rangle
=
-\sliver\half \!
\int_{-H}^0
   \rho\sliver gz\overline{(\xi_3^2)}
   _{\raisebox{1.3pt}{\scriptsize\antisliver$,33$}}
\Sliver dz
=
\half \!
\int_{-H}^0
   \rho\sliver g\sliver\overline{(\xi_3^2)}
   _{\raisebox{1.3pt}{\scriptsize\antisliver$,3$}}
\Sliver dz
=
\half\sliver \rho\sliver g\sliver
\overline{(\xi_3^2)}\big |_{z=0}
\;,
\label{eq:pot-en-in-pmom-law}
\end{equation}
which is the standard formula for the
surface-wave potential energy per unit area.

\vspace{-0.45cm}

\bibliographystyle{jfm}

\bibliography{refs}

\end{document}